\documentclass[twocolumn,trackchanges]{aastex7}
\usepackage{graphicx}
\usepackage{float}
\usepackage{siunitx}
\usepackage{multirow}
\usepackage{booktabs}
\usepackage{appendix}
\usepackage{chngcntr}

\newcommand{\ixpe}{\textit{IXPE}\xspace}

\newcommand{\xlcal}{\textit{XL-Calibur}\xspace}
\newcommand{\pogo}{\textit{PoGO+}\xspace}

\usepackage{xcolor}
\usepackage{gensymb}
\usepackage{xspace}
\usepackage{lipsum}
\usepackage{hyperref}
\usepackage{url}
\usepackage[symbol]{footmisc}

\newcommand{\nutilde}[0]{$\widetilde{\nu}$\xspace}
\newcommand{\nudottilde}[0]{$\dot{\widetilde{\nu}}$\xspace}
\newcommand{\alloff}{38}
\newcommand{\recovered}{36}
\newcommand{\nutildeeq}[1]{\widetilde{\nu}}
\newcommand{\nudottildeeq}[1]{\dot{\widetilde{\nu}}}

\begin{document}

\title{Refined Constraints on the Hard X-ray Polarization of the Crab Pulsar and Nebula Derived from an Extended \xlcal Dataset}

\correspondingauthor{*Jacob Casey (jacob.j.casey@unh.edu)}

\author[0000-0003-4433-1365]{Matthew G. Baring} \affiliation{Department of Physics and Astronomy -- MS 108, Rice University, 6100 Main Street, Houston, Texas 77251-1892, USA}\email{baring@rice.edu
} 

\author[0009-0009-3051-6570
]{Jacob Casey*
} \affiliation{Department of Physics and Astronomy and Space Science Center, University of New Hampshire, 8 College Road, Durham, NH 03824, USA
}\email{jacob.j.casey@unh.edu
}

\author[0009-0002-2488-5272]{Sohee Chun}  \affiliation{Physics Department and the McDonnell Center for the Space Sciences, Washington University in St. Louis, 1 Brookings Dr, Saint Louis, MO 63130, USA
}\email{sohee.chun@wustl.edu
}

\author[0000-0002-5250-2710
]{Ephraim Gau
} \affiliation{Physics Department and the McDonnell Center for the Space Sciences, Washington University in St. Louis, 1 Brookings Dr, Saint Louis, MO 63130, USA
}\email{ephraimgau@wustl.edu
}

\author[]{Tomohiro Hakamata}  \affiliation{Department of Earth and Space Science, The University of Osaka, 1-1 Machikaneyama-cho, Toyonaka, Osaka 560-0043, Japan
}\email{hakamata@ess.sci.osaka-u.ac.jp
}

\author[0000-0002-9705-7948
]{Kun Hu
} \affiliation{Physics Department and the McDonnell Center for the Space Sciences, Washington University in St. Louis, 1 Brookings Dr, Saint Louis, MO 63130, USA
}\email{hkun@wustl.edu
}

\author{Daiki Ishi
} \affiliation{Japan Aerospace Exploration Agency, Institute of Space and Astronautical Science, 3-1-1 Yoshino-dai, Chuo-ku, Sagamihara, Kanagawa 252-5210, Japan
}\email{ishi.daiki@jaxa.jp
}

\author[0000-0001-7477-0380
]{Fabian Kislat
} \affiliation{Department of Physics and Astronomy and Space Science Center, University of New Hampshire, 8 College Road, Durham, NH 03824, USA
}\email{fabian.kislat@unh.edu
}

\author[0000-0001-5191-9306]{Mózsi Kiss} \affiliation{KTH Royal Institute of Technology, Department of Physics, 106 91 Stockholm, Sweden}\affiliation{The Oskar Klein Centre for Cosmoparticle Physics, AlbaNova University Center, 106 91 Stockholm, Sweden}\email{mozsi@kth.se
}

\author[0000-0003-0441-4959
]{Merlin Kole
} \affiliation{Department of Physics and Astronomy and Space Science Center, University of New Hampshire, 8 College Road, Durham, NH 03824, USA
}\email{merlinkole@gmail.com
}

\author[0000-0002-1084-6507
]{Henric Krawczynski
} \affiliation{Physics Department, the McDonnell Center for the Space Sciences, and the Center for Quantum Leaps, Washington University in St. Louis, 1 Brookings Dr, Saint Louis, MO 63130, USA
}\email{krawcz@wustl.edu
}

\author[0009-0006-9014-2716
]{Haruki Kuramoto
} \affiliation{Department of Earth and Space Science, The University of Osaka, 1-1 Machikaneyama-cho, Toyonaka, Osaka 560-0043, Japan
}
\email{kuramoto@ess.sci.osaka-u.ac.jp
}

\author[0000-0002-5202-1642
]{Lindsey Lisalda
} \affiliation{Physics Department and the McDonnell Center for the Space Sciences, Washington University in St. Louis, 1 Brookings Dr, Saint Louis, MO 63130, USA
}\email{Lindsey.Lisalda@wustl.edu
}

\author[0009-0005-6415-1531
]{Bingkun Liu
} \affiliation{Physics Department and the McDonnell Center for the Space Sciences, Washington University in St. Louis, 1 Brookings Dr, Saint Louis, MO 63130, USA
}\email{bingkun@wustl.edu
}

\author[0000-0002-9099-5755
]{Yoshitomo Maeda} \affiliation{Japan Aerospace Exploration Agency, Institute of Space and Astronautical Science, 3-1-1 Yoshino-dai, Chuo-ku, Sagamihara, Kanagawa 252-5210, Japan
}\email{maeda.yoshitomo@jaxa.jp
}

\author{Hironori Matsumoto} \affiliation{Department of Earth and Space Science, The University of Osaka, 1-1 Machikaneyama-cho, Toyonaka, Osaka 560-0043, Japan
}\affiliation{Forefront Research Center, Graduate School of Science, The University of Osaka, 1-1 Machikaneyama, Toyonaka, Osaka 560-0043, Japan}\email{matumoto@ess.sci.osaka-u.ac.jp
}

\author[0009-0005-0818-7484
]{Shravan Vengalil Menon
} \affiliation{Physics Department and the McDonnell Center for the Space Sciences, Washington University in St. Louis, 1 Brookings Dr, Saint Louis, MO 63130, USA
}\email{s.vengalilmenon@wustl.edu
}

\author[0000-0002-6068-6337
]{Takuya Miyazawa
} \affiliation{Okinawa Institute of Science and Technology Graduate Universityy, Kunigami-gun, Okinawa, Japan
}\email{takuya.miyazawa@oist.jp
}

\author{Kaito Murakami} \affiliation{Department of Earth and Space Science, The University of Osaka, 1-1 Machikaneyama-cho, Toyonaka, Osaka 560-0043, Japan
}\email{murakami@ess.sci.osaka-u.ac.jp
}

\author[0000-0002-6054-3432]{Takashi Okajima} \affiliation{NASA Goddard Space Flight Center, Greenbelt, MD 20771, USA}\email{takashi.okajima@nasa.gov
}

\author[0000-0001-7011-7229
]{Mark Pearce} \affiliation{KTH Royal Institute of Technology, Department of Physics, 106 91 Stockholm, Sweden}\affiliation{The Oskar Klein Centre for Cosmoparticle Physics, AlbaNova University Center, 106 91 Stockholm, Sweden}\email{pearce@kth.se
}

\author[0000-0002-1452-4142]{Brian Rauch}  \affiliation{Physics Department and the McDonnell Center for the Space Sciences, Washington University in St. Louis, 1 Brookings Dr, Saint Louis, MO 63130, USA
}\email{brauch@physics.wustl.edu
}

\author{Kentaro Shirahama} \affiliation{Department of Earth and Space Science, The University of Osaka, 1-1 Machikaneyama-cho, Toyonaka, Osaka 560-0043, Japan
}
\email{shirahama@ess.sci.osaka-u.ac.jp
}

\author[0000-0003-0710-8893
]{Sean Spooner
} \affiliation{Department of Physics and Astronomy and Space Science Center, University of New Hampshire, 8 College Road, Durham, NH 03824, USA
}\email{sean.spooner@unh.edu
}

\author[0000-0001-6314-5897
]{Hiromitsu Takahashi
} \affiliation{Graduate School of Advanced Science and Engineering, Hiroshima University, 1-3-1 Kagamiyama, Higashi-Hiroshima, Hiroshima 739-8526, Japan
}\email{hirotaka@astro.hiroshima-u.ac.jp
}

\author[]{Sayana Takatsuka}  \affiliation{Department of Earth and Space Science, The University of Osaka, 1-1 Machikaneyama-cho, Toyonaka, Osaka 560-0043, Japan
}\email{takatsuka@ess.sci.osaka-u.ac.jp
}

\author[0000-0002-7962-4136
]{Yuusuke Uchida
} \affiliation{Japan Aerospace Exploration Agency, Institute of Space and Astronautical Science, 3-1-1 Yoshino-dai, Chuo-ku, Sagamihara, Kanagawa 252-5210, Japan
}\email{uchida.yuusuke99@jaxa.jp
}

\author[0000-0003-1471-2693
]{Varun
} \affiliation{KTH Royal Institute of Technology, Department of Physics, 106 91 Stockholm, Sweden}\affiliation{The Oskar Klein Centre for Cosmoparticle Physics, AlbaNova University Center, 106 91 Stockholm, Sweden}
\email{vvarun@kth.se
}

\author[0000-0002-5471-4709
]{Andrew Thomas West
} \affiliation{Physics Department and the McDonnell Center for the Space Sciences, Washington University in St. Louis, 1 Brookings Dr, Saint Louis, MO 63130, USA
}\email{Andrew.t.west@wustl.edu
}

\begin{abstract}
We present updated hard X-ray polarization measurements of the Crab pulsar and nebula obtained with the balloon-borne polarimeter \xlcal in the $\sim$\SIrange{19}{64}{keV} energy range. During the flight, intermittent GPS-failure; resulted in poorly constrained timing for \qty{\sim\alloff}{\percent} of the Crab dataset. By implementing a new phase-recovery method that reconstructs timing during extended GPS-off intervals, phase tag data is recovered for
$\sim$\qty{95}\% of the GPS-off dataset, increasing the precision of the phase-resolved analysis. Phase-information for the data is recovered by using the Crab pulsar, with its 33~ms period, as an external timing source. Using a Markov-Chain Monte-Carlo framework to jointly fit phase offsets and frequency derivatives, sufficient phase accuracy is achieved, across multiple periods without GPS for a phase-resolved analysis. This enables inclusion of nearly the full dataset in the polarization study. The polarization degree of the nebular emission is found to be \qty{27.7\pm4.9}{\percent} at a polarization angle of \ang{127.2}$\pm$\ang{5.1} confirming previous \xlcal results and remaining aligned with the Crab's spin axis, consistent with synchrotron emission from the inner nebula. Phase-resolved measurements show that the off-pulse and bridge intervals exhibit a strong polarization, while the pulsar peaks, although weakly constrained, remain in agreement with the softer-energy trends of \ixpe. These findings reinforce a scenario in which hard X-ray emission arises primarily in the nebular torus and wind regions. The successful recovery of precise phase tagging from GPS-off data demonstrates the capacity to use the pulsar as an external clock even in the case of sparsely populated data.

\end{abstract}



\section{Introduction} 

The Crab pulsar (PSR B0531+21) is one of the most studied sources in the X-ray sky, with its bright, stable, double‑peaked pulse profile comprising the main pulse (P1), inter‑pulse (P2), bridge (B) and off‑pulse (OP) phases. While the details of these structures can differ slightly in different energy ranges, such as the relative heights of P1 and P2, they serve as a precise phase reference across energy bands~\citep{1983ApJ...269..273W}.
Despite its status as a ``standard candle'' in the soft and hard X-ray bands, many questions, such as the origin of the high energy emission remain. 
Current models have the X-ray emission originating in the magnetosphere: a rapidly rotating, strongly magnetized plasma environment that accelerates particles to ultra-relativistic speeds. These particles emit radiation across the electromagnetic spectrum, modulated for an observer by the pulsar's rotation and shaped by complex electromagnetic field structures.  The polarization degree (PD) and polarization angle (PA) of the high energy emission, especially when resolved as a function of rotational phase, can distinguish between competing emission geometries, such as those predicted by outer-gap, slot-gap, and striped wind models, and test if the X-rays originate near the magnetic poles, in caustic magnetospheric regions, or from the wind termination shock ~\citep{Petri-2013-MNRAS,Harding-2017-ApJ,2019ApJ...871...12T,Kalapotharakos_2012,2025A&A...693A.152G}. 
In the hard X-ray regime, the photon flux from the nebula originates primarily from ultra-relativistic electrons in the inner torus and knot regions, where magnetic fields are predominantly toroidal and highly ordered \citep{Weisskopf.2000,Ng_2004}. 
Polarization measurements in the hard X-ray energy range therefore serve as diagnostics of magnetic-field geometry and turbulence in the post-shock flow. 

Recent \ixpe observations \citep{Bucciantini.2023,Wong.2024,ixpeCrab}, in the \SIrange{2}{8}{keV} energy range, have revealed spatial and energy resolved polarization across the nebula, with angles aligning to the pulsar spin axis at higher energies~\citep{ixpeCrab}. 
Extending these measurements to higher energies with \xlcal allows us to test whether the magnetic ordering persists in the population of freshly accelerated electrons responsible for the hard-X-ray emission~\citep{2025MNRAS.540L..34A}.
Comparison of PD and PA with those measured by \ixpe can constrain the energy dependence of the degree of field uniformity, and the efficiency of synchrotron cooling close to the termination shock \citep{2023PASJ...75.1298M,ixpeCrab,Abarr.2021}. In the soft to medium energy gamma-ray regime ($ > $\qty{100}{keV}), polarization measurements of the Crab have been obtained with \textit{INTEGRAL} and \textit{AstroSat}. \textit{INTEGRAL\slash IBIS} observations report a strong polarization in the \qty{200}~\unit{keV} to \qty{1}~\unit{MeV} band with the polarization aligned with the pulsar spin axis~\citep{2026arXiv260209886B}. Similarly, measurements with \textit{AstroSat} in the \SIrange{100}{380}{keV} range detect a high polarization fraction consistent with synchrotron emission from the inner nebula~\citep{2018NatAs...2...50V}. These measurements are broadly consistent with results from \pogo, which reported a PD of $20.9\pm5.0 \%$ from the Crab in the \SIrange{18}{160}{keV} band, comparable to the energy range probed by \xlcal~\citep{Chauvin.2017,Chauvin.2018c7}. Polarimetry in the \xlcal, $\sim$\SIrange{19}{64}{keV}, energy range can further probe synchrotron emission to offer crucial insight into the particle acceleration processes and emissions closest to the pulsar.

\begin{figure*}[t!]
\centering
\includegraphics[width=0.75\linewidth, trim={0 0 0 0}, clip = true]{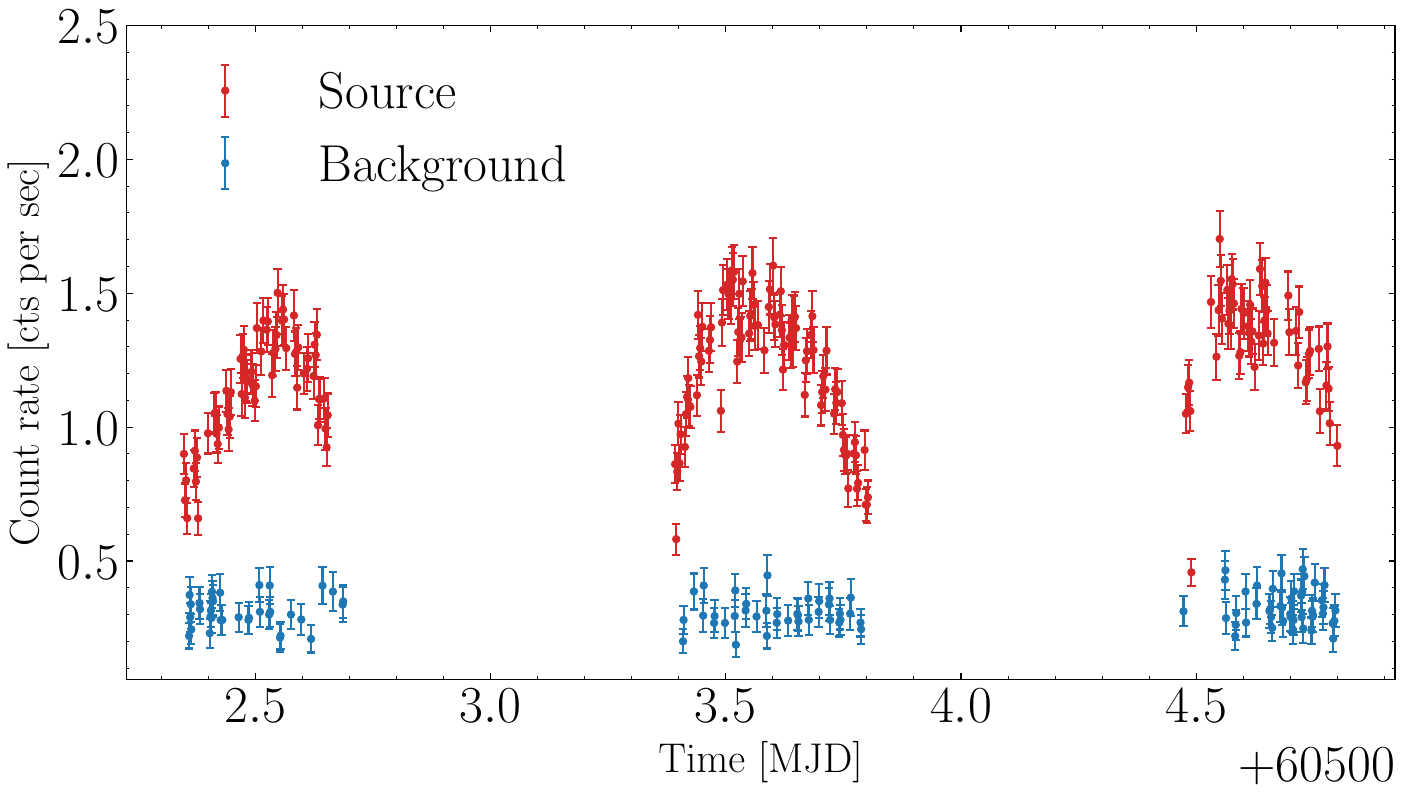}
\caption{
On-source (red) and off-source (blue) count rates during the three Crab observations (Days~2--4). 
Each cluster of points corresponds to a single observation day. Observations of Cyg X$-$1 occur during gaps between observations when the Crab is out of view, see \citet{cyg} for more details on these observations.
The modulation of the on-source rate is caused by changes in pointing elevation, with the source's motion in the sky, and atmospheric attenuation, while the background rate remains stable, demonstrating consistent detector performance.
}
\label{fig:ratecurve}
\end{figure*}

\begin{figure*}[t!]
    \centering
    \includegraphics[width=1\linewidth]{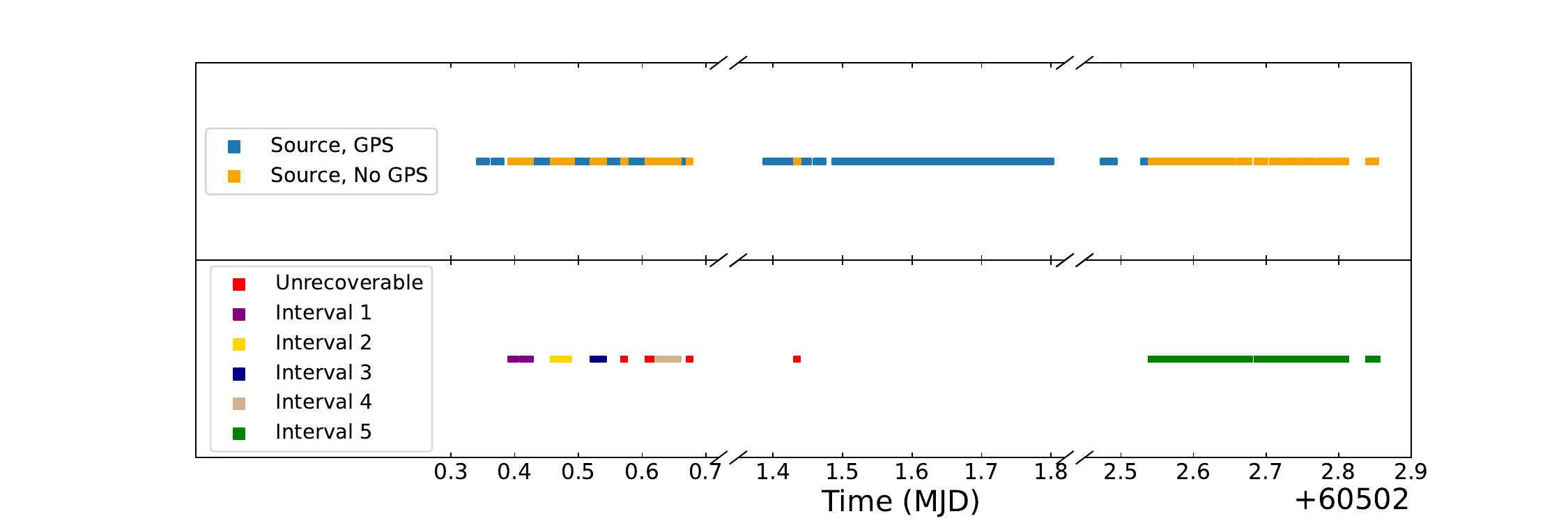}
    \caption{Time segmentation of the Crab observation showing GPS-on and GPS-off intervals. The x-axis is truncated to only depict observations of the Crab. While the GPS problem persisted during the Cyg X$-$1 observations, the same data loss did not occur due to the lower required precision for absolute timing for the source. Source events with valid GPS timestamps are shown on the top panel in blue, while those lacking GPS timestamps are shown in orange. Labeled intervals (Intervals 1–5 in the bottom panel) denote the segments used for template construction and timing recovery. The short ``Unrecoverable'' intervals (red) represent brief GPS-off data excluded from the analysis. This segmentation defines the structure of the timing correction pipeline, with each segment requiring an individual phase offset.}

    \label{fig:FlagBreakdown}
\end{figure*}

The balloon-borne mission  \xlcal  offers a high signal-to-background platform for hard X-ray polarimetry in the $\sim$\SIrange{19}{64}{\keV} energy range, where the Crab is one of the brightest astrophysical sources~\citep{2025MNRAS.540L..34A}. \xlcal  is a Compton polarimeter that achieves polarization sensitivity by measuring the azimuthal scattering angles of incoming photons~\citep{Abarr.2021}. During its 2024 flight, the instrument observed the Crab nebula and pulsar for a total source exposure of \qty{\sim 50}{ksec} and an additional \qty{\sim 17}{ksec} of interspersed background pointing, to extract both phase-integrated and phase-resolved polarization signatures. Additional observations of Cyg X$-$1 occurred while the Crab was below the horizon. Initial results of the Crab and Cyg X$-$1  are presented in~\citet{2025MNRAS.540L..34A,cyg} respectively. However, due to intermittent failures of the GPS system on the payload, approximately \qty{\alloff}{\percent} of the total on-source Crab events lacked valid GPS event timestamps, where an internal oscillator is used for time-tagging instead, which lacked sufficient stability for a phase-resolved study.
As a result, this data originally had to be excluded from the phase-resolved analysis~\citep{2025MNRAS.540L..34A}.

The focus of this paper is to detail a method to reclaim the ``GPS-off'' fraction of the dataset, using on-board clocks and the Crab pulse profile derived directly from the GPS‑tagged portion of the same XL‑Calibur observation. The methods described here are based on techniques well-established in the radio pulsar community, where template matching and timing residual analysis are routinely employed to track spin-down parameters, glitches, and binary motion~\citep[e.g.,][]{2010MNRAS.402.1027H, 2024A&A...687A.154W}. In the context of radio-astronomy, the cause of mismatched phase tagging is typically due to uncertainty in the pulse period rather than clock timing, but the symptom of inaccurate event phase tagging remains. This work adapts and applies similar principles to the challenge of phase tagging X-ray events in the absence of GPS data. This method allows us to expand the \xlcal dataset by \qty{\sim\recovered}{\percent}, phase-resolving \num{\sim24000} more photons. 
As a result, the constraints on the polarization degree improve by a factor of \num{\sim 1.2}.

In order to address the phase tagging challenge we (i) describe the extraction and characterization of the Crab pulse template from the GPS‑tagged data in Section~\ref{sec: Methods}; (ii) present the phase-correction algorithm and quantify it's phase‑recovery performance (Section \ref{Sec:Template Extraction}); and finally (iii) apply the method to the intervals lacking proper timing, demonstrating an accurate reconstruction of the phase‑folded light curve (Sections~\ref{sec:CrabDay4}~and~\ref{sec:CrabDay2}). The updated, phase-tagged data is used to constrain the PD and PA across P1 and P2 with increased sensitivity.The phase tagging allows us to enhance not only the dataset for the pulsed emission, but also the dataset for the nebular emission. Because XL-Calibur does not spatially resolve the pulsar from the nebula, the nebular contribution is isolated using the rotational phase. In this work we refer to the off-pulse interval as the phase region where the pulsed emission is minimal and the observed flux is dominated by nebular emission (\SIrange{0.57}{0.87}{revs}). 
The improved constraints and discuss the implications are presented in Section~\ref{sec:PolAnalysis}.

\section{Data Processing}
Data for this analysis was collected during the 2024  \xlcal balloon flight. Over the course of the mission, three science observations were conducted of the Crab pulsar and nebula. The three science observation days (MJD 60502–60504), referred to here as Crab Days 2–4, constitute the primary  \xlcal  Crab dataset, corresponding to OBSIDs: 202401000[2-4] on the \textit{HEASARC} data archive. During Days 2 and 4, thermal management issues intermittently disrupted the GPS signal used to maintain time synchronization with ground-based communication, essential for timing stability across observations. When this occurred, data continued to be collected using an internal oscillator referenced to the clock of the flight computer. This system provides the same timing granularity as the GPS-based clock, but only provided absolute time accuracy of the order of 1 second and was found to drift on the O($10^{-5}$) level. This enabled sufficiently accurate time tagging to determine if events are from on-source or off-source pointing. See Figure~\ref{fig:ratecurve} for the relative source and background count rates as a function of time for the Crab observations. 

The backup internal pulse-per-second (PPS), supplied by a \qty{100}{MHz} crystal oscillator in the instrument's Truss CPU, was originally intended for bench-testing detectors. The internal oscillator time provides a relative, not absolute, timing source. It is not directly linked to any continuously running clock, instead it is synchronized to the computer's CPU time within one second precision.  The internally generated PPS has a different phase offset from the CPU clock every time it is enabled. This means that each time the GPS status changes, a phase offset occurs in the backup clock's reference frame, originating from the re-syncing of the clocks and making corrections across multiple GPS-off intervals more difficult. Additionally, the duration between each PPS pulse is sensitive to flight conditions, such as temperature, that result in slow drift in the timing stability from a ``true PPS''.

During Day 2 of the observation the GPS functioned with intermittent failures, resulting in a fragmented dataset where some periods had accurate timing and some did not. On Day 3, the GPS-operated normally with only one brief outage. During Day 4, however, the GPS had a total failure early in the day and remained unrecoverable after this point. The result is a single, continuous GPS-off segment. The separation of the dataset into GPS-on and GPS-off intervals is illustrated in the top row of Figure~\ref{fig:FlagBreakdown}. 

Following the identification of GPS-on and GPS-off intervals, events are classified as on-source or off-source. Accurate polarization reconstruction requires a stable definition of the scattering geometry and optical axis across the three observation days. Small variations in pointing or optical alignment can introduce systematic biases in the reconstructed azimuthal scatter angle distribution if left uncorrected. In order to account for these effects, average pointing offsets are measured for each observation using the imaging CZT detector. The average pointing offsets are used to calculate azimuthal scatter angles. Details about these calculations are found in Appendix~\ref{App A}.

With this geometry established, polarization observables are constructed from the event level azimuthal scatter angles using the standard Stokes formalism for Compton polarimeters~\citep{KISLAT201545}. Event level Stokes parameters are accumulated over the relevant time, energy, and phase selections, and background subtraction is performed using off-source pointings.

\section{Methods}
\label{sec: Methods}

In order to recover phase information from GPS-off intervals, a template-based timing correction procedure was developed. First, a template pulse profile is generated from GPS-on data. This template is then used as a reference for correcting timing intervals without GPS timing. Each interval, defined by uninterrupted use of the internal oscillator requires a unique phase offset, generated by the re-syncing of the on-board clock upon reestablishment of GPS signal. For improved statistics, Intervals 1--4 on Day 2 are grouped together, sharing a frequency and frequency derivative. Interval 5 on Day 4 is corrected independently from the other Intervals. Each segment was fit to a model given by:
\begin{equation}
\label{eq:PhaseDef}
\phi = \phi_i +\nutildeeq~(\Delta{t})+ \frac{1}{2}\nudottildeeq~(\Delta{t})^2
\end{equation}
A new frequency \nutilde, and frequency derivative \nudottilde for the pulsar, is introduced, defined relative to our clock's time rather than GPS time. Additionally, $\Delta{t}$ is the time since the mean event time for the day of observation, $\Delta{t} = t - \bar{t}$. This model will allow effects such as clock drift to be incorporated into the definitions of \nutilde and \nudottilde. A phase offset $\phi_i$, is allowed, relative to the GPS-on reference interval. These values will be defined relative to the mean time of the GPS-off interval they describe.
The subscript \textit{i} denotes the correction interval of each phase offset $\phi_i$.
These parameters are optimized by maximizing the agreement between the folded light curve from the GPS-off data and a template derived from the GPS-on intervals as discussed in more detail in Sections~\ref{sec:CrabDay4}~and~\ref{sec:CrabDay2}.

This approach effectively re-aligns the internal oscillator to the pulsar's true rotational phase using only observational data, without requiring external timing references. It enables high-fidelity reconstruction of the Crab's pulse profile even in the absence of valid GPS timing, thereby extending the usable fraction of the dataset for phase-resolved analysis. In order to recover accurate phase tagging of events, it is assumed that any clock drift can be accounted for by selecting the appropriate values for the observed pulsar frequency \nutilde and its derivative \nudottilde, in the clock's timing reference frame.

\subsection{Template Extraction}
\label{Sec:Template Extraction}
In order to recover a pulse profile that most closely matches what is expected from the corrected GPS-off intervals, a template is constructed directly from the subset of data with valid GPS-on events. Deriving the template directly from  \xlcal  data ensures consistency with the instrument response, avoids cross-calibration issues with external observatories, and enables precise phase recovery in the data segments that lack proper timing.

\begin{figure}
    \centering
    \includegraphics[width=0.9\linewidth]{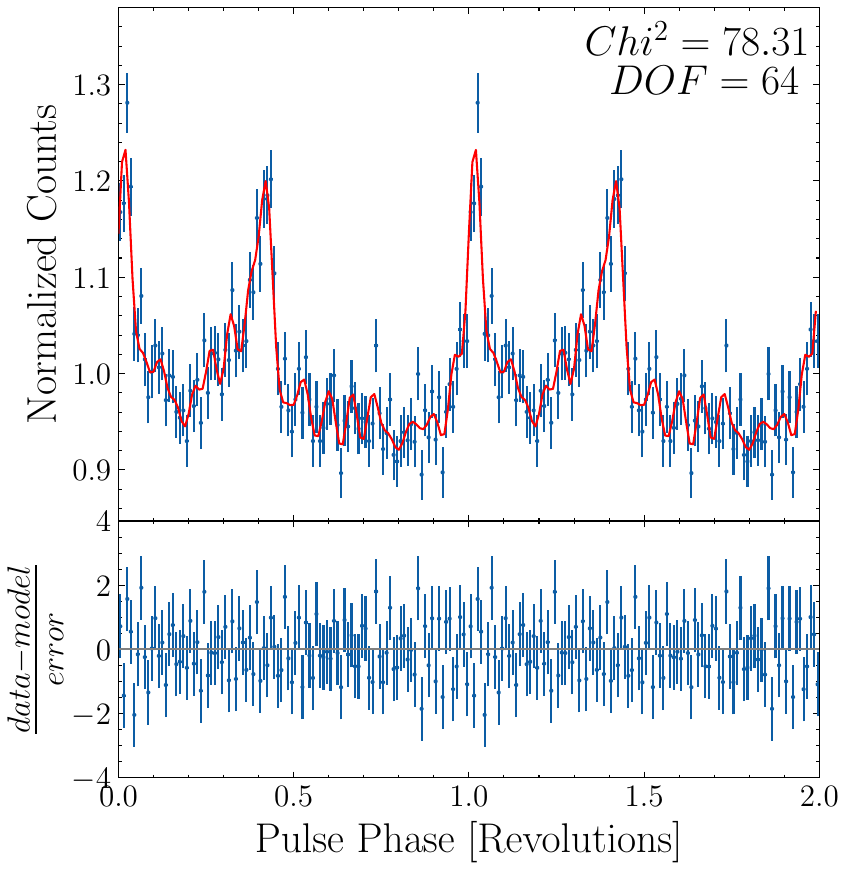}
    \caption{ Folded light curve of the Crab pulsar derived from GPS-on data shown in blue, overlaid with the best fit $18$-term Fourier template in red. The pulse profile is shown over two full rotations for clarity. The agreement between the data and the smooth model confirms the adequacy of the harmonic expansion and supports it's use as a reference template for phase recovery in GPS-off intervals.}
    \label{fig:FourierSeriesGPS-on}
\end{figure}

The GPS-on events were barycentered, and folded modulo the Crab's instantaneous rotation period, as determined by the Jodrell Bank ephemeris~\citep{10.1093/mnras/stu2118}\footnote[2]{\url{http://www.jb.man.ac.uk/pulsar/crab.html}} dated 2024 June 15, which is the closest preceding ephemeris to the flight, to construct a pulse profile. As in~\citet{2025MNRAS.540L..34A}, the peak of P1, as measured by \textit{XL-Calibur}, is defined as phase zero. The peak location is found by fitting a narrow Gaussian to P1 in the GPS-on dataset, as a result, phase definitions differ from radio-based definitions by a constant offset due to a ``lag'' of $\sim$\qty{0.01}{revs} between the X-ray and radio peaks~\citep{2004ApJ...605L.129R}. \textit{XL-Calibur} uses a phase-integrated background determined from off-source pointing. The template extraction and phase correction are performed on non-background subtracted data to reduce statistical uncertainty in the count rates. The background is subtracted for the final pulse profile generation and subsequent polarization analysis. Here a negligible change in the Crab's pulse profile over the 3 days of \textit{XL-Calibur} observation is assumed, allowing use of an observation-integrated pulse profile.

Relying on direct histograms from GPS-on data could amplify noise and introduce binning artifacts due to limited statistics. Instead, a folded light curve was modeled by a truncated Fourier series. The number of harmonics in this model was determined using the $H$-test, a non-parametric statistical method designed to assess the significance of periodic signals in sparse or noisy data~\citep{1989A&A...221..180D}.  The $H$-test returns an optimal number of harmonics $M$ that balances signal recovery with overfitting risk by evaluating the improvement in fit, as successive harmonics are added to a Fourier expansion of the data.

Applying the $H$-test to the GPS-on data yielded an optimal harmonic number of $M = 18$, which was used to perform a discrete Fourier transform (DFT) of the folded pulse profile. The result is a smooth, noise-suppressed analytic representation of the Crab pulsar's X-ray light curve in the \qtyrange{19}{64}{\keV} band. The Fourier coefficients $a_k$ and $b_k$ define the pulse template function
\begin{equation}
\label{eq:FourierSeries}
\widetilde{f}(\phi) = \sum_{k = 1}^{M} \bigl[a_k \cos({2\pi{k}\phi}) + b_k \sin({2\pi{k}\phi})\bigr],
\end{equation}
where $\phi \in [0, 1)$ is the rotational phase. The template is normalized such that $\int_0^1 \widetilde{f}(\phi) \, d\phi = 1$, facilitating direct comparison with normalized, observed count-rate variations in the GPS-off data.

Figure~\ref{fig:FourierSeriesGPS-on} depicts the folded light curve derived from the GPS-on interval overlaid with the fitted Fourier template. The agreement between the binned data and the smooth harmonic model illustrates the ability of the 18-term Fourier expansion to capture the essential structure of the pulse profile, including the sharp peaks of P1 and P2, an increasing flux in the bridge emission, and the off-pulse baseline. Using this approach, rather than simply aligning peak maxima, enables a more robust phase recovery, particularly in cases where estimates exist for the pulsar frequency and its derivative. By incorporating additional structure in the pulse profile, such as the detailed peak shapes and the gradual increase in the bridge region, this method improves phase precision beyond what is achievable with peak matching alone.

\subsection{Crab Day 4}
\label{sec:CrabDay4}

The longest GPS-off segment in the \textit{XL-Calibur} Crab dataset occurred on Day 4  (MJD 60504) of the flight, spanning a contiguous half-day interval in which absolute timestamps were unavailable. The lack of GPS mode switching, occurring during this period allows the entire GPS-off portion of Day 4, Interval 5, to be treated as a single timing unit, requiring just three parameters to recover its absolute phase alignment.


\begin{figure}
    \centering
    \includegraphics[width=0.8\linewidth]{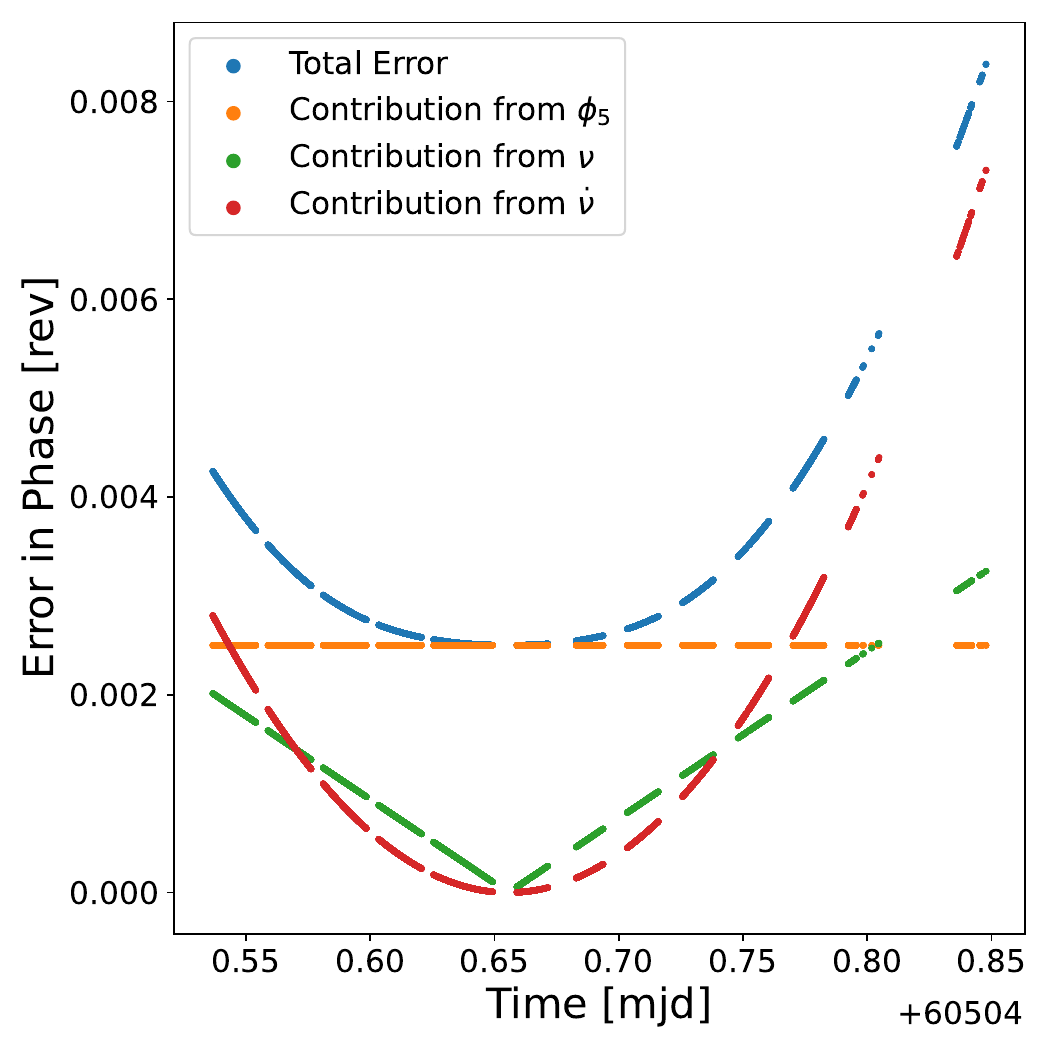}
    \caption{Propagation of phase uncertainties from the MCMC posterior distributions into total phase error across the Day 4 interval. Contributions from each parameter offset: $\phi_5$ (orange), \nutilde (green), and \nudottilde (red), are shown individually, along with the combined total phase uncertainty (blue). The maximum error at any point remains well below the width of phase intervals used for phase-resolved polarization analysis, supporting the stability of the timing solution. This is critical for the polarization analysis in Section~\ref{sec:PolAnalysis}, where bin-to-bin migration of events could bias the final result. }
    \label{fig:Day4ErrorValues}
\end{figure}

\begin{figure}
    \centering
    \includegraphics[width=0.9\linewidth]{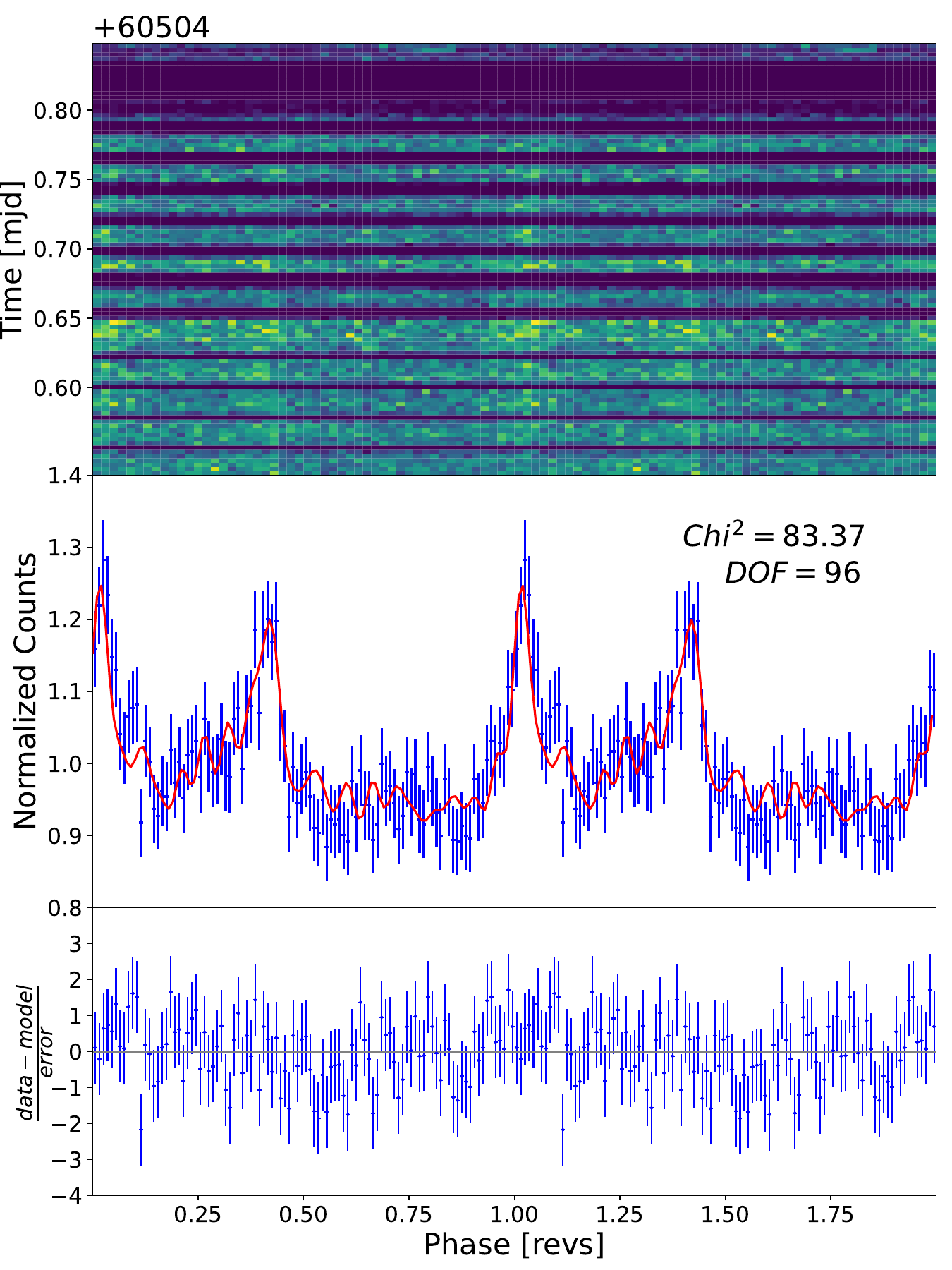}
    \caption{The top panel depicts the time- and phase-resolved pulse profile, with phase on the horizontal and time on the vertical axis. The color scale corresponds to the number of events in a given bin. Brighter regions depict bins with higher count rates. Clear vertical lines are visible at $\phi \sim 1$ and $\sim 1.3$ due to the pulsar peaks. The middle panel shows the reconstructed, time-integrated   pulse profile for the Day 4 GPS-off interval (blue), Interval 5, folded using the timing parameters recovered from MCMC and overlaid with the template profile (red) derived from GPS-on data. The middle panel has been normalized to allow direct comparison to the model; the bottom panel shows residuals (data minus model), with a reduced $\chi^2$ of 83.37/96, with a $\chi^2$ probability of \num{\sim 0.82}, confirming the quality of the timing correction and the reliability of the phase recovery.}
    \label{fig:Day4PulseP}
\end{figure}


Initial values of $\phi_5$,~\nutilde, and \nudottilde, defined relative to the mean GPS-off photon detection time, were estimated through an optimization that minimized the $\chi^2$ value of the folded light curve compared to the pulse template derived in Section~\ref{Sec:Template Extraction}. This is accomplished by allowing one parameter to vary at a time, iteratively, while being conscious of potential local minima by scanning large regions of the parameter space. The resultant first approximation of $(\phi_5,\nutildeeq~, \nudottildeeq/)$ are consistent with visually aligned peaks in the folded profile, but further inspection indicated a complex parameter space, making it difficult to quantify the uncertainty of the fit parameters. For this reason, the minimizer results were then used to initialize a Markov Chain Monte Carlo (MCMC) analysis using the \textsc{emcee} package~\citep{2013PASP..125..306F}, which samples the posterior probability distribution for each parameter, assuming Gaussian likelihoods based on the $\chi^2$ distributions. During the MCMC, limits were placed on allowed values of $\phi_5$ to avoid a local minima in the case of one peak aligning with the incorrect peak of the template (i.e. P1 of the template aligning with P2 of the data). For day 4, the MCMC samples 2 million parameter configurations, while 15 million are used for Day 2 due to a more complex function topology. The mode of the distributions is used for folding the light curve instead of the mean value found by the MCMC. For singly peaked distributions they are typically compatible, but this convention breaks ambiguity in the case of bi-modality. In cases of bi-modality, it is common for the mean to occur somewhere between the two peaks of the distribution, where fit quality is generally less optimal. The mode of the distribution however, by definition occurs in the peak. The underlying distributions of the MCMC for Day 4 are shown in Figure~\ref{fig:Day4MCMC}.

In the GPS-off segments on Day 4, the MCMC posterior distributions for the phase offset $\phi_5$ exhibits mild bimodality. In this case the secondary peaks are separated by approximately one bin width of the pulse profile and result from limited statistics. The modes are drawn from the underlying MCMC posterior in the full $N$-dimensional parameter space, where $N$ is the number of free parameters (3 for Day 4 and 6 for Day 2). This approach avoids relying on one-dimensional histogram projections or two-dimensional slices of the posterior. Working directly with the full posterior reduces biases introduced by binning and more accurately reflects the structure of the sampled parameter space.

To account for timing uncertainties in the phase reconstruction, the posterior spreads in $\phi_5$, \nutilde, and \nudottilde were propagated to phase error as a function of time. This propagation is shown in Figure~\ref{fig:Day4ErrorValues}, where the contributions due to each parameter are plotted separately, along with the total phase uncertainty. The larger of the MCMC error bounds for each parameter were used to ensure conservative estimates. The resulting phase uncertainty is compatible with the pulse profile bin width of \qty{0.01}{revs}, even a maximum shift resulting from Figure~\ref{fig:Day4ErrorValues} would not shift an event further than the adjacent bin. Additionally, the phase uncertainty is much smaller than \SIrange{0.2}{0.3}{revs}, corresponding to the phase intervals used in a phase-resolved analysis (P1, P2, bridge, and off-pulse intervals), confirming the stability of the solution over the entire GPS-off interval. Phase interval definitions for the phase-resolved analysis can be found in Table \ref{tab:polValues}.

 \begin{table*}
     \caption{
     Table of correction values given by the mode of an MCMC distributions shown in Appendix~\ref{App B}. $T_{ref}$ is given as the point of mean flux for events during the day. All other correction parameters are defined relative to this time. The forward propagated frequency from the Jodrell Bank ephemeris is \qty{29.559248}{\Hz} for Day 2~\citep{10.1093/mnras/stu2118}.}

     \centering
     \hspace*{-2.2cm}
     \begin{tabular}{lccccc}
         \toprule
         \multicolumn{2}{c}{Correction Interval} & $T_\mathit{ref}\;[\mathrm{mjd}]$ & $\phi_i\;[\mathrm{revs}]$ & $\nutildeeq\;[\frac{\mathrm{revs}}{\mathrm{s}}]$ & $\nudottildeeq\;[10^{-9}\,\frac{\mathrm{revs}}{\mathrm{s}^{-2}}]$\\
         \midrule
        Day 2 & 1 &$60502.521322$& $0.250$ & $29.5597467$ & $2.89$  \\

         & 2 &\vdots & $-0.151$ &  \vdots &  \vdots \\

         & 3 & \vdots&$ 0.758$ &\vdots   & \vdots  \\

         & 4& \vdots& $-0.143$ &\vdots  &\vdots  \\
         \midrule
         Day 4 & 5 & $60504.655607$& $0.4855$ & $29.55914044$ & $1.67$ \\
         \bottomrule
     \end{tabular}
     \label{tab:CorrectionValues}
 \end{table*}
\subsection{Crab Day 2}
\label{sec:CrabDay2}

Unlike Day 4, where GPS timing was unavailable for a single, continuous interval, Day 2 of the \textit{XL-Calibur} Crab observation (MJD 60502) was marred by intermittent periods without GPS, resulting in multiple short segments of phase-less data broken up by GPS-on intervals. This presents a distinct challenge: each GPS-off segment will have its own relative phase offset, while it's assume that the corrected pulsar's spin frequency (\nutilde) and spin down rate (\nudottilde) remain constant throughout each observation. Variations in the PPS's timing stability are relatively small over a day of observation, allowing Intervals 1-4 to be fit simultaneously, and be validated by performing independent fits of each region, where frequency results were compatible within statistics.   
To recover phase coherence across these distinct intervals, a fit of all recoverable Day 2 GPS-off data was performed with a single \nutilde and \nudottilde, allowing each interval an independent phase offset $\phi_i$.  It was found that timing was recoverable for four of the seven GPS-off intervals during Day 2. Shorter GPS-off periods had insufficient statistics to constrain the phase offset.

As with Day 4, initial estimates of parameters $\phi_i$, \nutilde, and \nudottilde were obtained by minimizing the $\chi^2$ between each segment's folded profile and the template described in Section~\ref{Sec:Template Extraction}. An MCMC is initialized with these initial values. The MCMC chains produced posterior distributions for $\phi_i$, \nutilde,and ~\nudottilde, (Figure~\ref{fig:Day2MCMC}, diagonal panels), with well-constrained modes and quantified uncertainties. When compared with Day 4, the MCMC posterior distributions for the Day 2 phase offsets, $\phi_i$, exhibited stronger bimodality as a consequence of less exposure in each interval. The dominant peak mode in each distribution was adopted as the optimal value, indicated by the central lines in Figure~\ref{fig:Day2MCMC}, to ensure consistent phase alignment. The modes are drawn from the unbinned, underlying MCMC result that make up each histogram on the diagonal.

\begin{figure}
    \centering
    \includegraphics[width=0.45\textwidth]{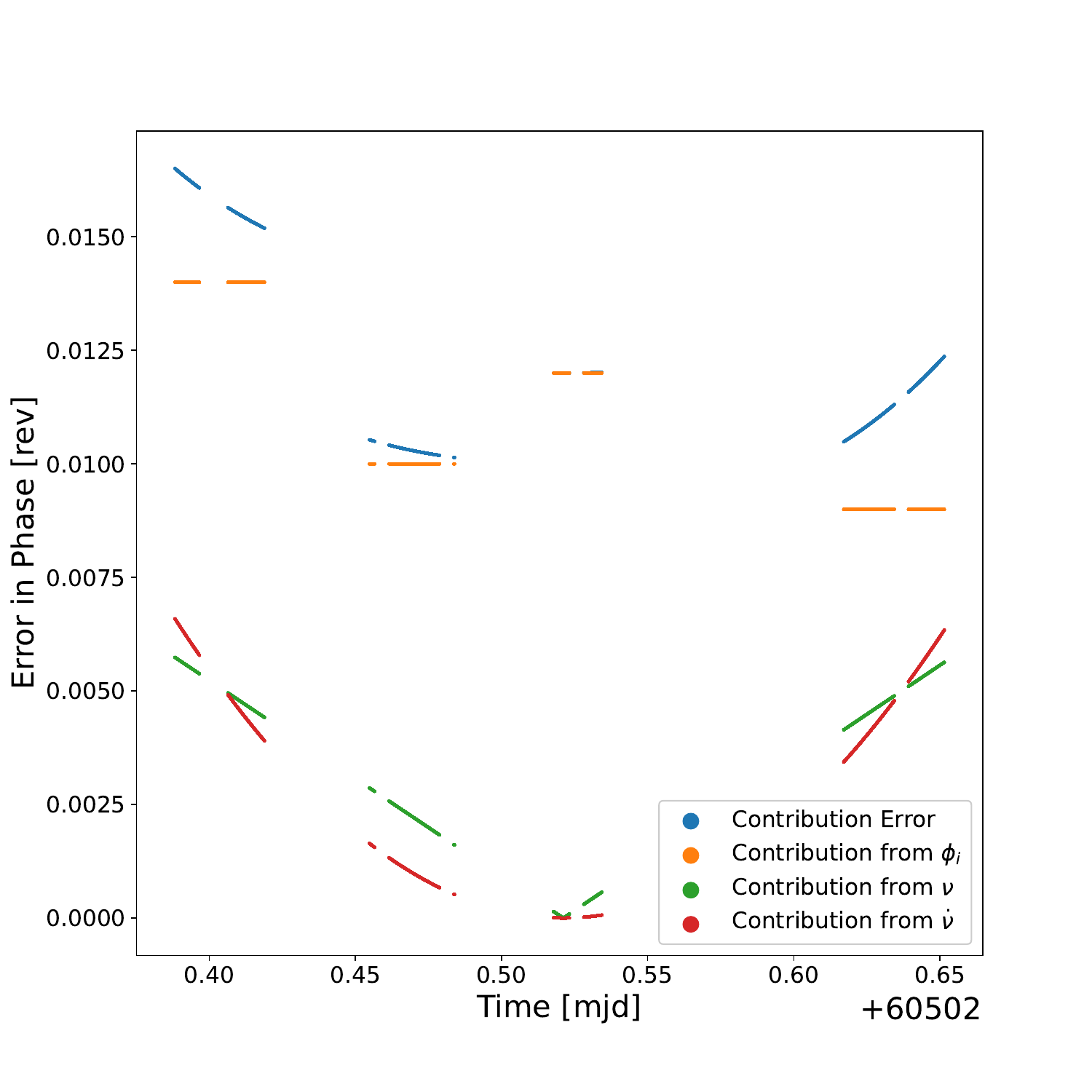}
    \caption{ Similar to Figure~\ref{fig:Day4ErrorValues}, propagation of phase uncertainties from the
    MCMC posterior distributions into total phase error across
    the Day 2 interval. Contributions from each parameter: $\phi_i$ (orange), \nutilde   (green), and \nudottilde (red) are shown individually, along with the combined total phase uncertainty
    (blue). The maximum error at any recovered time is well below the phase interval definitions used for phase-resolved polarization analysis.}
    \label{fig:Day2Errors}
\end{figure}

Uncertainties in $\phi_i$ were propagated to timing error estimates across the corresponding GPS-off intervals to ensure consistent phase alignment. These propagated errors are shown in  Figure~\ref{fig:Day2Errors} for each segment. In all cases, the maximum error at any point remains well below the width of phase intervals used for P1, P2, bridge, and off-pulse analysis in Section~\ref{sec:PolAnalysis}, ensuring the validity of phase-resolved analysis (e.g., P1 vs. P2 polarization). Note that the uncertainty contribution from \nutilde and \nudottilde for Interval 3 is nearly zero, this is because the frequency and frequency derivative are defined relative to a point within this Interval, the center of the observation.


\begin{figure}[hbt!]
    \centering
    \includegraphics[width=0.5\textwidth]{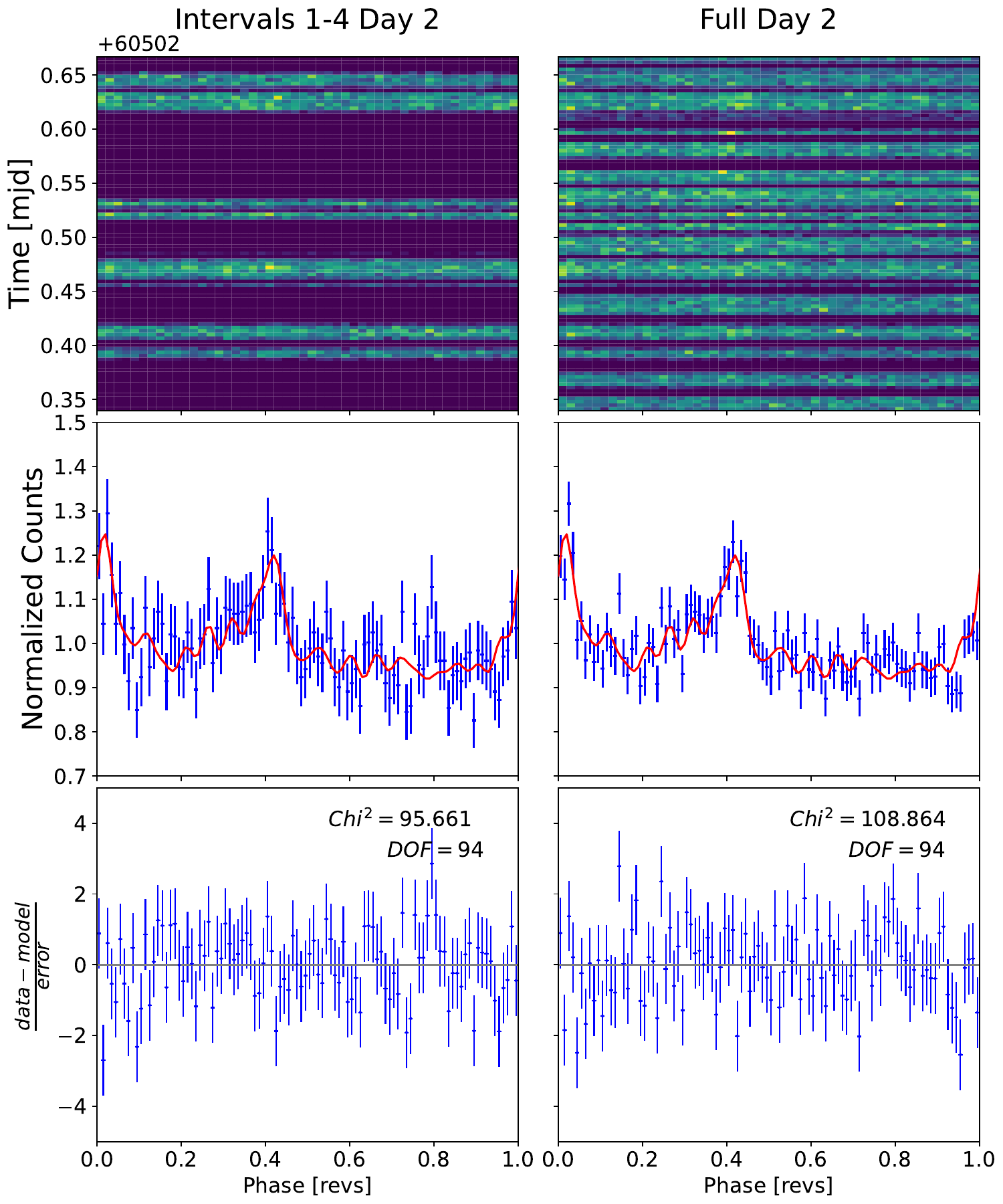}
    \caption{Recovered phase-folded light curves for Day 2 of the Crab observation. The top row shows the event density in phase versus time for the Day 2 GPS-off segments after phase correction (left) and for the full Day 2 interval (right). Only GPS-off intervals are used in fitting; the full Day 2 result is shown only for comparison. The middle panels display the normalized folded light curves, with the red curve representing the fitted Fourier template. The bottom panels show residuals (data - model)/error for each case. The inclusion of GPS-off segments using the recovered phase offsets preserves the pulse morphology, demonstrating the effectiveness of the timing-correction procedure for fragmented GPS-off intervals.}
    \label{fig:PulseProfileD2}
\end{figure}

The GPS-off data segments were then folded using the recovered $\phi$, \nutilde and \nudottilde, then plotted alongside the Fourier template for validation (Figure~\ref{fig:PulseProfileD2}, middle panels). The folded profiles consistently aligned with the expected pulse morphology, providing strong evidence that the phase reconstruction was successful across multiple periods without GPS.

\section{Polarization Analysis}
\label{sec:PolAnalysis}

Utilizing the phase information successfully recovered from the fragmented GPS-off intervals, updated polarization results are calculated for the Crab pulsar using the full reconstructed dataset. The addition of the previously phase-less \qty{\recovered}{\percent} of on-source events improved the statistical precision of the phase-resolved Bayesian analysis of Stokes parameters.
\begin{figure}
    \centering
    \includegraphics[width=0.95\linewidth]{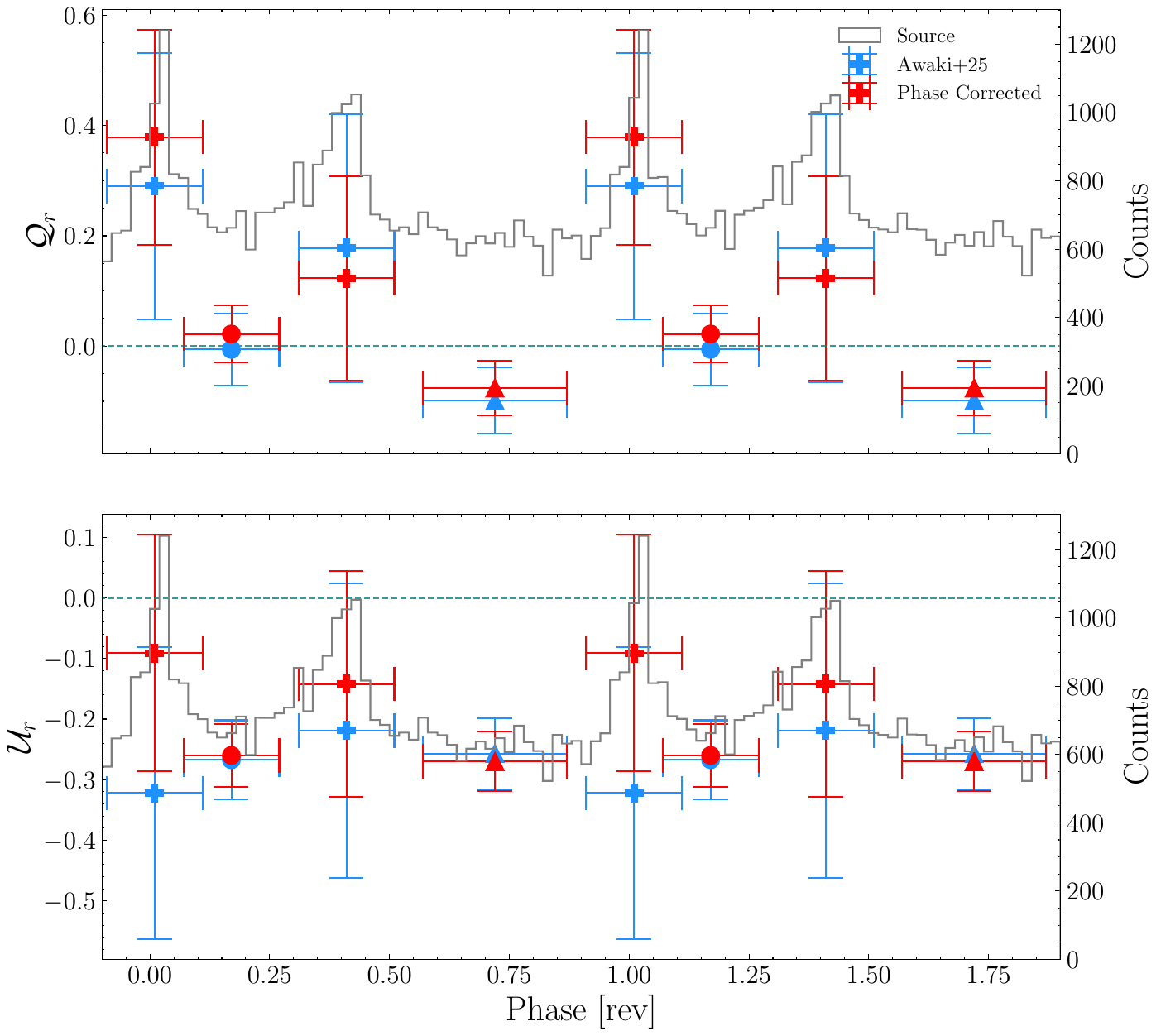}
    \caption{Phase-resolved Stokes parameters $Q_r$ (top) and $U_r$ (bottom) for the Crab pulsar in the \qtyrange{19}{64}{\keV} band, shown over two full rotations for clarity. Results are shown for four phase intervals: main pulse (P1), bridge, inter-pulse (P2), and off-pulse. Results from the full phase-corrected dataset, including GPS-on intervals, are shown in red, while only {GPS-on} results, previously reported in~\citep{2025MNRAS.540L..34A}, are in blue. Error bars denote 1$\sigma$ uncertainties on the Stokes parameters. A background-subtracted pulse profile and horizontal error bars are plotted to clarify the phase intervals of each point. The nebula (triangle) and bridge (circle) are background-subtracted using data collected during off-source pointing. P1 and P2 (cross) are background-subtracted using the nebula (triangle) polarization result.}
    \label{fig:CrabQU}
\end{figure}

\begin{figure}
    \centering
    \includegraphics[width=1\linewidth,trim={0 4mm 0 20mm},clip=true]{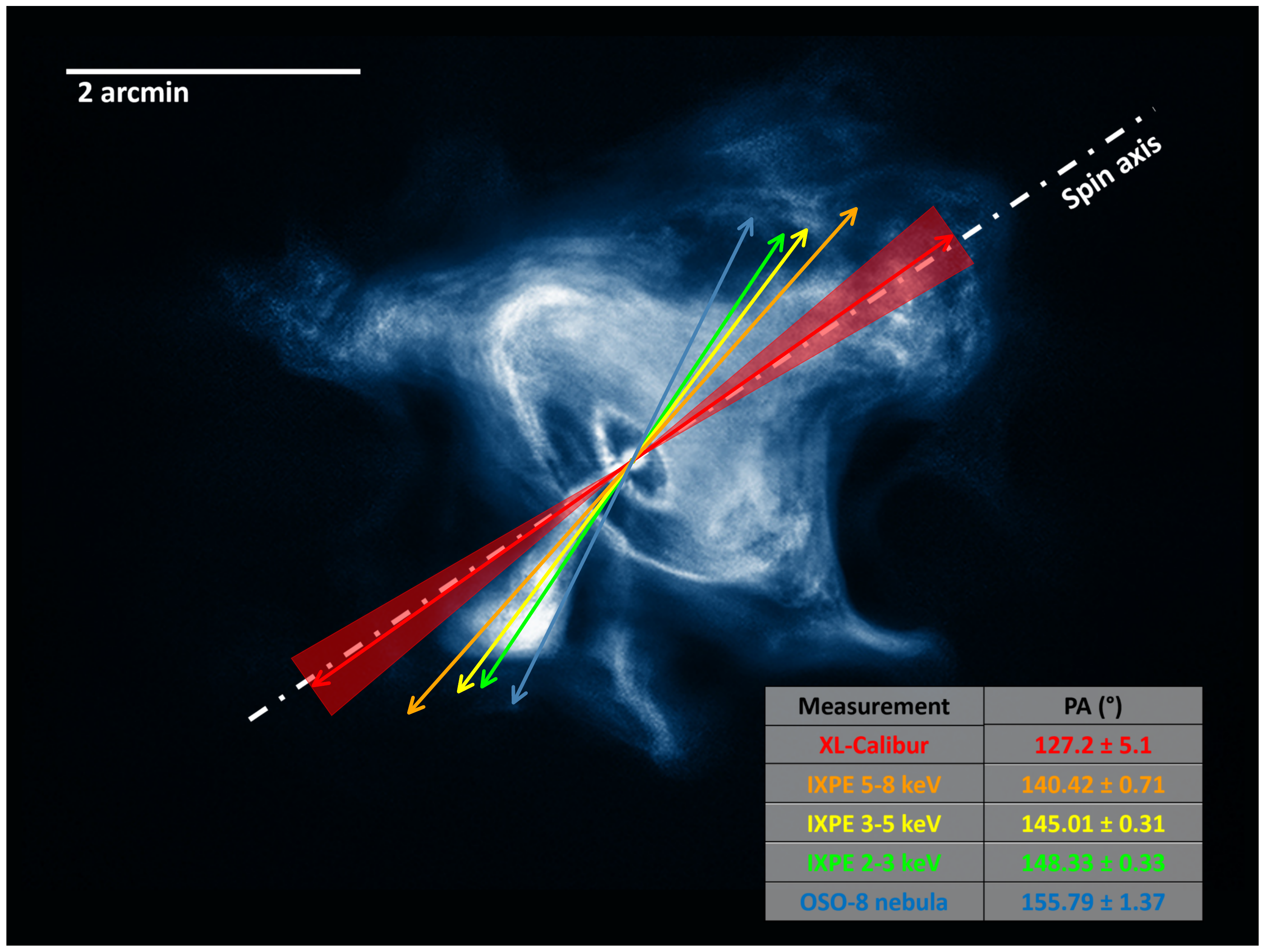}
    \caption{ Energy-resolved polarization measurements of the Crab Nebula emission from \ixpe and OSO$-$8, reported in \citet{ixpeCrab} and \citet{1978ApJ...220L.117W} respectively, and compared with the expanded \xlcal results. Data from \xlcal follows the trend where as energy increases, the nebular polarization approaches the pulsar's spin axis. The arrow length is proportional to the PD in each energy band. This figure is updated from \citet{2025MNRAS.540L..34A}.}
    \label{fig:Nebula}
\end{figure}

Results from a Stokes analysis of the updated dataset are shown in Figure~\ref{fig:CrabQU} in comparison with GPS-on data previously reported in~\citep{2025MNRAS.540L..34A}. The polarization signal is extracted in four phase intervals: P1, P2, bridge, and off-pulse. For the polarization analysis, the non-background-subtracted off-pulse polarization is used as the background for P1 and P2. This is done to isolate the polarization contribution of the pulsar in a way that minimizes the uncertainty. The increased sensitivity is particularly evident in the pulsar peaks due to improved photon counts in both source and background (off-pulse) data subsets. No improvement is made through this analysis to the off-source observations that serve as background for the off-pulse and bridge regions, this is because these background observations are phase-integrated, so there is no benefit from the higher timing accuracy.

\begin{figure*}
    \centering
    \includegraphics[width=0.49\linewidth]{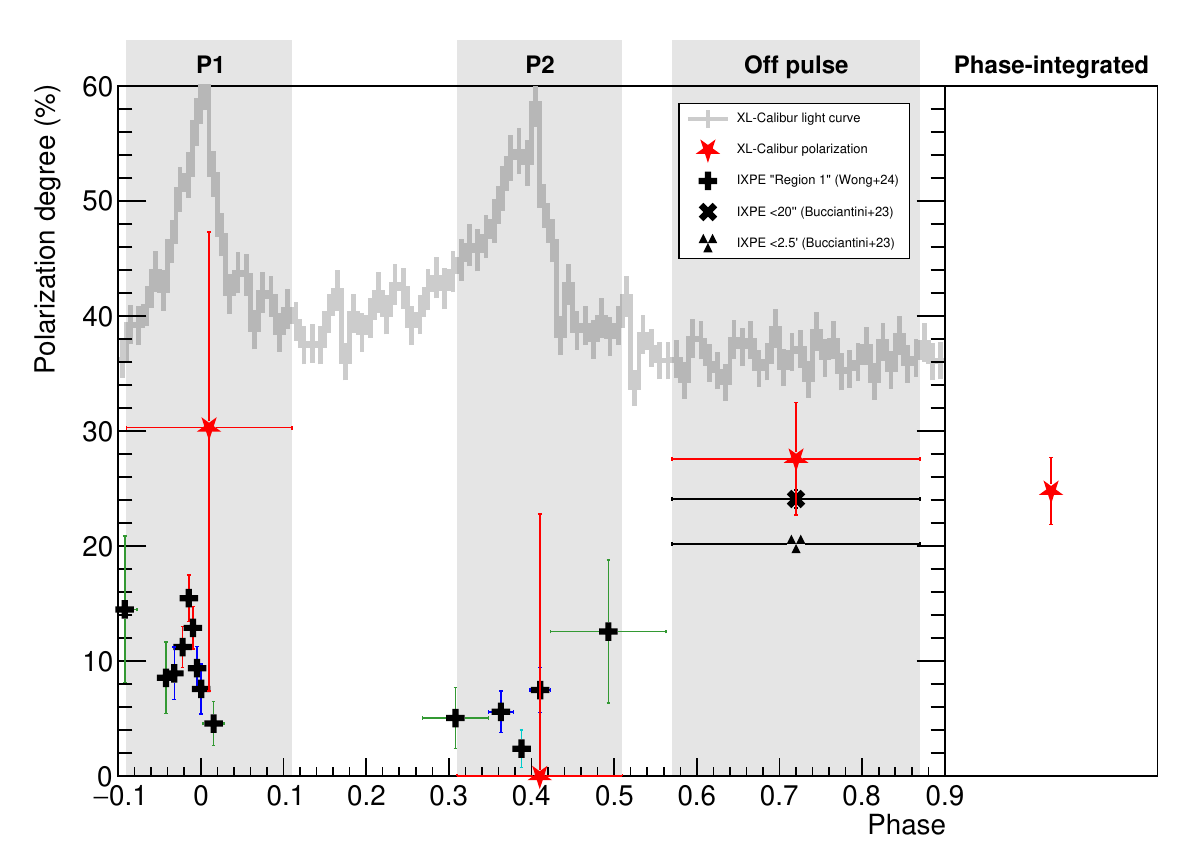}
    \includegraphics[width=0.49\linewidth]{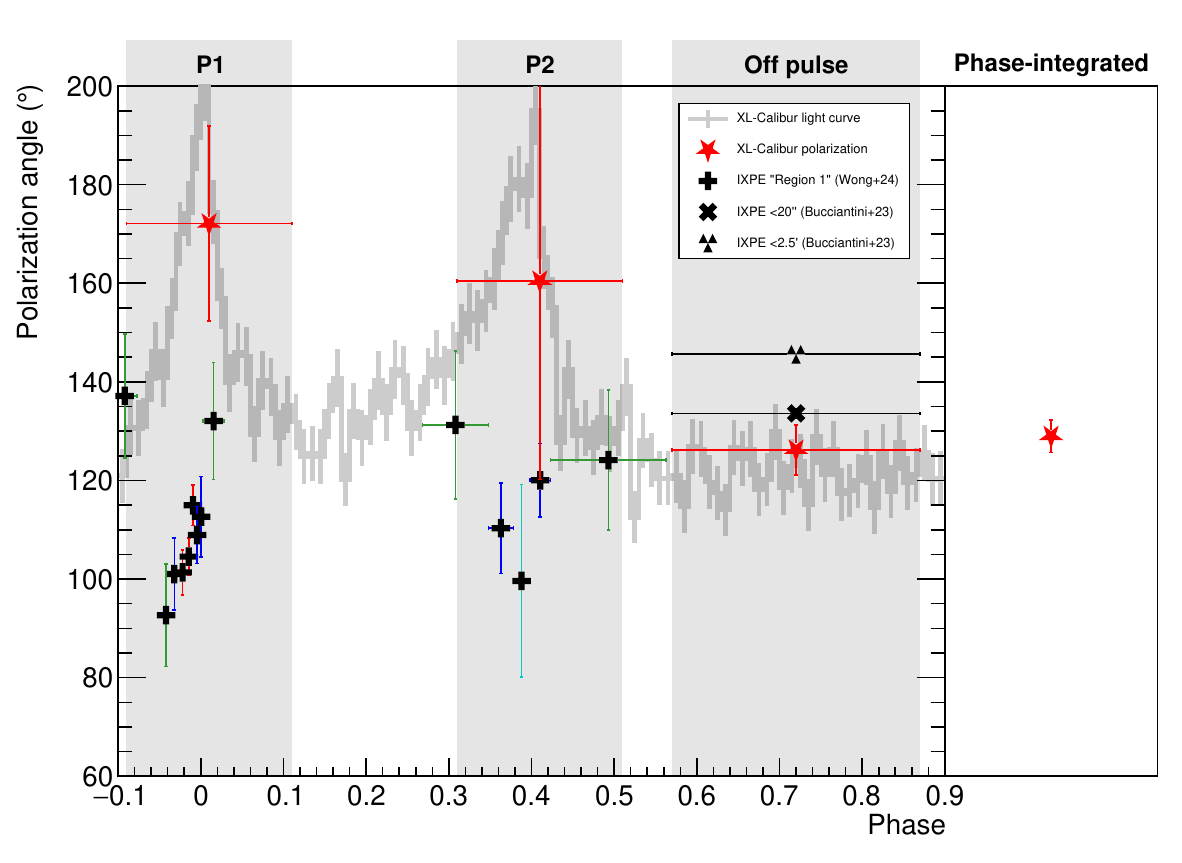}
    \caption{Phase-resolved polarization degree (left) and polarization angle (right) for the Crab pulsar in the \qtyrange{19}{64}{\keV} band, as measured by XL Calibur (red stars), using updated phase tagging when compared to Figure 2 of \citet{2025MNRAS.540L..34A}. Shaded regions indicate the main pulse (P1), inter-pulse (P2), and off-pulse intervals used in the analysis. The background gray curve shows the pulse profile for reference. Comparison points are drawn from some previous \textit{IXPE} observations,~\citet{Bucciantini.2023} and~\citet{2024ApJ...973..172W}. The \textit{IXPE} error-bar colors indicate different levels of significance ($>5\sigma, >3\sigma, >1.9\sigma, <1.9\sigma$ for red, blue, green and turquoise, respectively). The  \textit{XL-Calibur}  results include both GPS-tagged and phase-recovered data, leading to improved constraints on PD and PA across all bins in comparison to results from~\citet{2025MNRAS.540L..34A}.}
    
    \label{fig:PD-PAphase-resolved}
\end{figure*}
The resulting phase-resolved PD and PA are shown in Figure~\ref{fig:PD-PAphase-resolved}. For the bridge and off-pulse phase bins, the MDP was calculated using 
\begin{equation}
\label{eq:MDP-offpulse}
\hspace{-3mm}
\mathrm{MDP} = \frac{4.29}{\mu_{100}(\frac{N_{i}}{\Delta\phi T_{on}} - \frac{N_{off}}{T_{off}})}\sqrt{\frac{N_{i}}{(\Delta\phi T_{on})^2} + \frac{N_{off}}{(T_{off})^2}}
\end{equation}
    where $N_{off}$, and $T_{off}$ correspond to background counts and exposure respectively, while $\Delta\phi$ is the phase bin width and $N_{i}$ is the on-source count rate within the phase bin~\citep{KISLAT201545,Weisskopf2006}. 
For the P1 and P2 bins, the nebula serving as the background, the MDP equation differs slightly:
\begin{equation}
\label{eq:MDP-pulse}
\mathrm{MDP} = \frac{4.29}{\mu_{100}(\frac{N_i}{\Delta\phi} - \frac{N_n}{\Delta\phi_n})}\sqrt{\frac{N_i}{(\Delta\phi)^2} + \frac{N_n}{(\Delta\phi_n)^2}},
\end{equation}
where $\Delta{\phi_n}$ is the phase bin width of the nebula and $N_n$ is the number of counts in that bin. Nebula rates without background subtraction are used for $N_n$, to reduce the uncertainty in the effective background.
The nearly energy-independent, modulation factor, $\mu_{100}$, for  \textit{XL-Calibur}  is reported in~\citet{2025MNRAS.540L..34A}, and takes a value of ${0.424}\pm{0.001}$. This value is calculated using the average spectral shape of the Crab, for more details on the calculation see~\citet{2024APh...15802944A}. The background for the off-pulse and bridge phase ranges is defined by the periods of off-source pointing.

A full summary of these results is presented in Table~\ref{tab:polValues} and Figure~\ref{fig:PD-PAphase-resolved}. Across all phase ranges, the calculated MDPs are consistently lower in the full dataset, reinforcing the value of recovering GPS-off data. 

The main pulse (P1) region in particular exhibits the largest adjustment in polarization results, but remains statistically consistent with the results presented in~\citet{2025MNRAS.540L..34A}. The modest reduction in uncertainty was expected due to the signal rate in P1 of \qty{\sim 1}counts per second compared to a Nebula/Background rate of \qty{\sim 0.3}counts per second. The additional exposure improves the MDP from 73\% to 61\%. This improvement is not sufficient for a robust detection, demonstrating the need for further data collection to enable a more definitive measurement. As seen in P1, fluctuations in polarization results can be expected when measuring below MDP level~\citep[e.g.,][]{2018A&A...615A..54M,2025arXiv250404775L}. The inter-pulse (P2) region remains consistent with zero polarization, and while the MDP improves slightly (from 74\% to 59\%), the recovered PD remains below detection threshold and the marginalized posterior still peaks at zero. 

\begin{table*}
    \caption{
    The background-subtracted, normalized Stokes $Q_r, U_r$ results, and the marginalized PD and PA values from the Bayesian analysis for the  \textit{XL-Calibur}  observation. Following~\citet{2025MNRAS.540L..34A}, the prior that has been used is uniform in polar coordinates (PD,PA). Results presented here are updated from Table 2 of ~\citet{2025MNRAS.540L..34A} with the additional phase-recovered data from Intervals 1-5, the updated results are well within statistical agreement. The phase-integrated results are identical to those previously published. Definitions of each phase range are drawn from~\citet{Bucciantini.2023} and~\citet{2024ApJ...973..172W}. The maximum a posteriori (MAP) estimate is the mode (most probable value) of the two-dimensional Bayesian posterior.  PD and PA result from marginalizing over the posterior; uncertainties stated are credible regions corresponding to $1\sigma$ Gaussian probability content. Background subtraction is performed using data from the phase-integrated off-source pointing, without requirements on the GPS status, except in the case of P1 and P2 where the off-pulse, prior to background subtraction, serves as background. For P2 only an upper limit for the polarization is set.}

    \centering
    \hspace*{-2.2cm}
    \begin{tabular}{cccccccc}
        \toprule
        Observation & Phase Range & \textbf{$Q_r$} & \textbf{$U_r$} & MAP (\%, $^{\circ}$ )&PA ($^{\circ}$) & PD ($\%$) & $MDP_{99}$ ($\%$)\\
        \midrule
        
       Phase Integrated & 0-1 & $-0.045 \pm 0.029$ & $-0.249 \pm 0.029$& $(25.3,129.8)$ & $129.8 \pm 3.2$ & $25.3^{+2.8}_{-2.9}$ & $8.7$ \\

        Off-pulse & \numrange{0.57}{0.87} & $-0.076 \pm 0.049$ & $-0.270 \pm 0.049$& $(28.0, 127.1)$&$127.1 \pm 5.1$ & $27.7\pm{4.9}$ & $15$  \\

        Bridge &\numrange{0.07}{0.27} & $0.022 \pm 0.052$ &$ -0.260 \pm 0.052$&$(26.1,137.4)$ &$137.4 \pm 5.9$  & $25.7^{+5.2}_{-5.3}$ & $16$ \\

        P1 (``Main pulse") & \numrange{-0.09}{0.11} &$ 0.379 \pm 0.195$ & $-0.091 \pm 0.196$& $(39,173.2)$ &$173.2 \pm 18.8$ & $32.4^{+17.9}_{-22.7}$ & $60$  \\

        P2 (``Inter-pulse") & \numrange{0.31}{0.51} & $0.123 \pm 0.185$ & $-0.142 \pm 0.185$&$(18.8,155.5)$ &$155.5 \pm 38$ & $0^{+23.8}_{-0} $& $56$ \\

        \bottomrule
    \end{tabular}
    \label{tab:polValues}
\end{table*}

\section{Discussion and Interpretation}

The primary result of this work is the successful recovery of rotational phase information during extended GPS-off intervals in the \xlcal Crab observation. Using the Crab pulsar itself as an external timing reference and applying a template-based phase-recovery procedure, approximately \qty{\recovered}\% of the on-source exposure that was previously excluded from phase-resolved analysis in \citet{2025MNRAS.540L..34A} was reclaimed. This increases the effective data volume without introducing further systematic effects in the reconstructed pulse profiles or polarization measurements.

As an additional validation of the phase-recovery procedure, the same timing correction pipeline was applied to contemporaneous observations of Cyg X$-$1 and to background pointings obtained during the same flight. 
In both cases, the recovered phase-folded light curves show no statistically significant periodicity, and the fitted parameters are unconstrained, as expected for steady sources. 
This null-result provides strong evidence that the successful phase recovery observed for the Crab GPS-off intervals arises from genuine pulsar periodicity rather than from algorithmic bias or instrumental systematics.

Compared to the earlier results derived from the smaller dataset in~\citet{2025MNRAS.540L..34A}, the recovered dataset provides increased statistical significance for the polarization of all phase bins. The largest improvements being for the pulsar peaks (P1, P2) due to better precision in both the ``source'' and in the off-pulse which serves as the background. The polarization results from this analysis are summarized in Table \ref{tab:polValues}.


Despite these improvements, the polarization information of pulsar peaks remain only weakly constrained. As such, the interpretation of the results is unchanged from the results previously presented in~\citet{2025MNRAS.540L..34A}. The marginalized posterior of the expanded dataset for P1 peaks at a PD of $(32.4^{+17.9}_{-22.7})\%$, while P2's marginalized posterior peaks at zero. Upper limits for the PD of P1 and P2 at the 99\% Confidence Level are 78.75\% and 56.35\% respectively. This pattern mirrors the trend found in \ixpe data of higher polarization in P1 than in P2 and of rapid PA rotation across P1~\citep{2024ApJ...973..172W}. The absence of a strong, stable polarization signal in the hard-X-ray band reinforces the interpretation that emission near the light-cylinder is dominated by rapidly varying field orientations that depolarize the phase-averaged signal. Models invoking emission from caustic or outer-gap regions of the magnetosphere, where projected magnetic-field directions vary rapidly with phase, remain consistent with our results.
For example, the comparatively modest PD in P1 and low PD in P2 are approximately commensurate with the range of model predictions from synchrotron emission zones proximate to the light cylinder or in the striped wind just outside it \citep{Petri-2013-MNRAS,Harding-2017-ApJ}.
In contrast, inner-gap models predicting much higher PDs are disfavored ~\citep{2019ApJ...871...12T,Kalapotharakos_2012,2025A&A...693A.152G}.  

The phase-integrated and off-pulse polarization, dominated by nebular emission, remains consistent with previous hard-X-ray results from \textit{PoGO+}, \textit{OSO-8}, and \xlcal~\citep{pogo, pogo+,1978ApJ...220L.117W,2025MNRAS.540L..34A} and with \textit{IXPE}'s soft X-ray maps~\citep{Bucciantini.2023,Wong.2024}, see Figure \ref{fig:Nebula}. 
With the inclusion of the phase-recovered GPS-off data, the nebular polarization measurement reaches a statistical significance of $\sim5.7\sigma$, compared to $\sim4.5\sigma$ in ~\citet{2025MNRAS.540L..34A}, while remaining fully consistent with earlier results.
The measured off-pulse PD of $27.7 \pm 4.9\%$ and PA of $127.15^\circ\pm5.1^\circ$ align with the projected spin axis of the pulsar \citep{Weisskopf.2000}, reaffirming that the hard X-ray emission arises primarily in the toroidal magnetic structure of the inner nebula rather than in the outer, more turbulent filaments, and following the trend seen by other missions (\textit{OSO-8}, \textit{IXPE}, \pogo, \textit{AstroSat}, \textit{INTEGRAL}) that as energy increases, the PA gets closer to the spin axis.

The success of the phase recovery demonstrates that pulsar timing information can be reconstructed reliably using the pulsar itself as an external clock, even in the absence of continuous GPS synchronization. The recovered phase uncertainties remain well below the widths of phase intervals used in Section~\ref{sec:PolAnalysis}, ensuring that bin-to-bin migration does not bias the polarization measurements. 

Unlike traditional pulsar timing applications that rely on long-term ephemerides or high signal-to-noise profiles on a per-period basis, this method operates effectively in low-count regimes by exploiting the cumulative structure of the pulse profile over extended intervals. This makes it particularly well suited to hard X-ray and gamma-ray balloon missions and satellite platforms where telemetry, power, or thermal constraints can compromise timing stability.

More generally, this work demonstrates that scientifically valuable phase-resolved analysis need not be discarded solely due to partial timing failures. By treating each uninterrupted internal-clock interval independently and fitting for phase offsets and effective pulsar spin parameters within the instrument's timing reference frame, it is possible to recover phase coherence across complex observing sequences. This approach may enable the recovery of archival datasets previously deemed unusable for phase-resolved studies and can be incorporated into the design of future missions as a contingency strategy for timing anomalies.

Looking forward, further improvement in phase-resolved hard X-ray polarimetry of the Crab will require either multi-week balloon flights in the Northern Hemisphere or an orbital mission employing a similar detector concept. Improved constraints on the pulsar polarization would benefit from spatial separation of the pulsar and nebular emission, although achieving imaging capability in the hard X-ray band presents significant instrumental challenges. Concepts such as imaging polarimeters could, in principle, disentangle these components, but require careful tradeoffs between modulation factor, background, and angular resolution. The phase-reconstruction method presented here demonstrates that usable pulsar timing can be recovered despite repeated GPS dropouts and system resets. In hindsight, maintaining a stable free-running clock with the GPS disabled would likely have preserved phase coherence more effectively than repeated re-initialization of the GPS timing system. Future instruments should therefore prioritize an onboard clock with sufficient intrinsic stability, such as a rubidium-based oscillator, to eliminate this failure mode, while retaining correction techniques such as those presented here as a safeguard for non-ideal operating conditions.

\section{Acknowledgments}
This work builds on results previously reported in \citet{2025MNRAS.540L..34A}, with an expanded dataset.
{\it XL-Calibur} is a joint mission supported by NASA, JAXA, and the Swedish National Space Agency {\it Rymdstyrelsen}. 
We acknowledge NASA support under grant 80NSSC24K0205.
KTH authors are supported by the Swedish National Space Agency (2022-00178 and 2024-00248). MP also acknowledges funding from the Swedish Research Council {\it Vetenskapsrådet} (2021-05128).
The Washington University in St. Louis group acknowledges additional NASA support through the grants 80NSSC20K0329, 80NSSC21K1817, 80NSSC22K1291, 80NSSC22K1883, 80NSSC23K1041, and 80NSSC24K1178, as well as funding from the McDonnell Center for the Space Sciences at Washington University in St. Louis. 
The University of New Hampshire group acknowledges additional NASA support through the grants 80NSSC24K0636, 80NSSC24K1762, and 80NSSC25K8001.
The Japanese Society for the Promotion of Science (JSPS) has supported this work through KAKENHI Grant Numbers 19H01908, 19H05609, 20H00175 (HM), 20H00178 (HM), 21K13946 (YU), 22H01277 (YM), 23H00117, and 23H00128 (HM).

\clearpage
\appendix
\section{Appendix A}
\setcounter{figure}{0}
\setcounter{table}{0}
\counterwithin{table}{section}
\counterwithin{figure}{section}
\label{App A}
\begin{figure}[H]

\centering
\includegraphics[width=0.3\linewidth, trim={2mm 0 0 8mm}, clip = true]{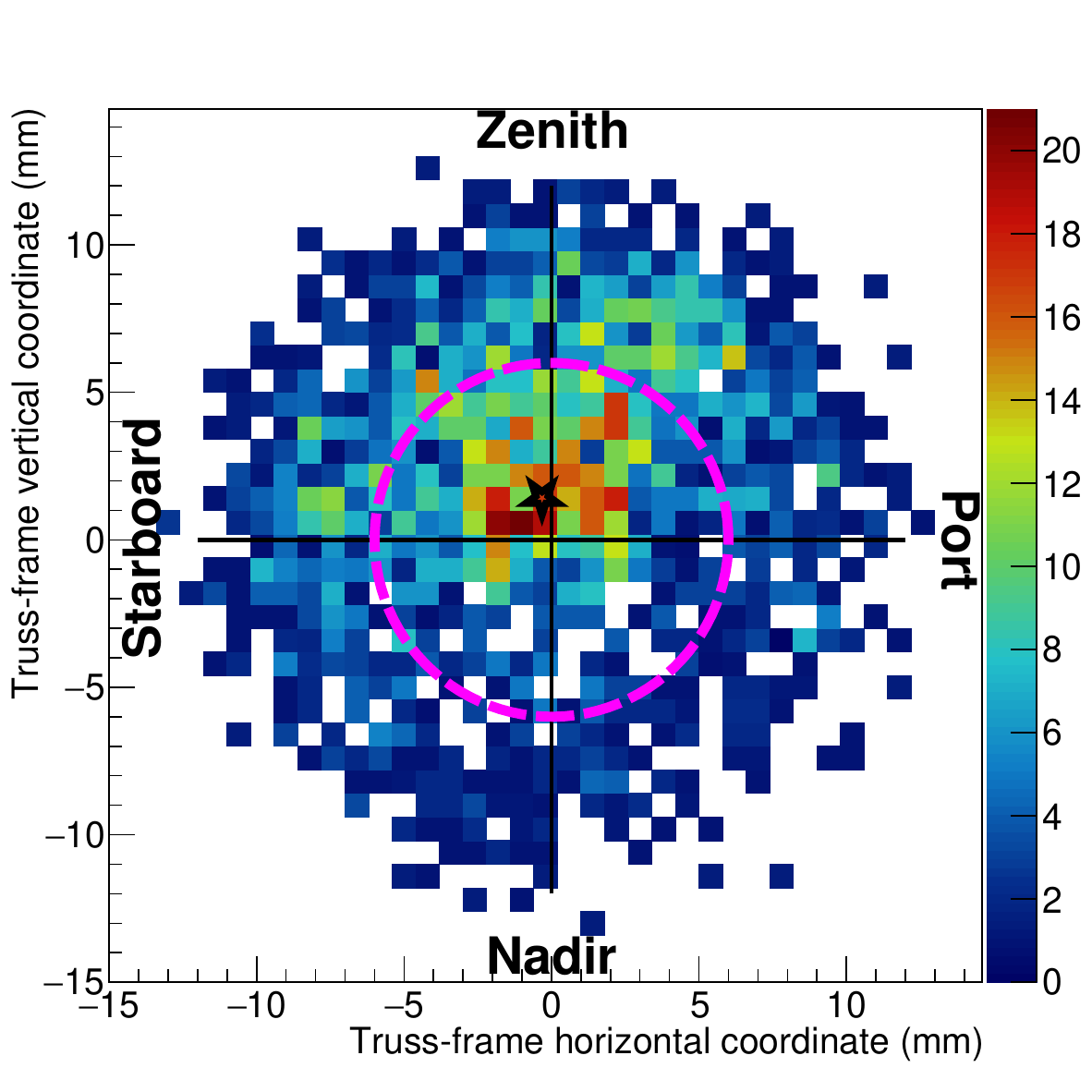}
\includegraphics[width=0.3\linewidth, trim={2mm 0 0 8mm}, clip = true]{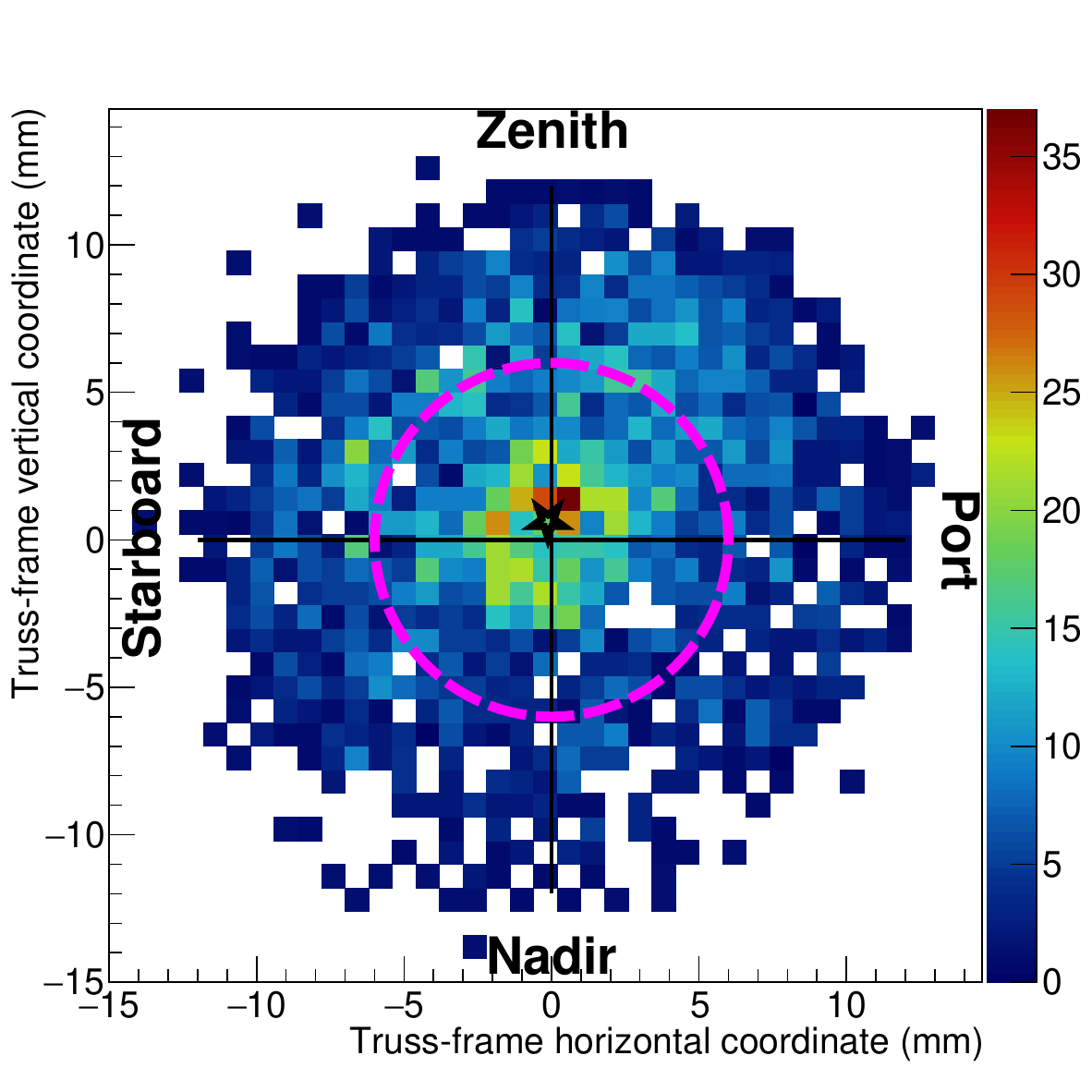}
\includegraphics[width=0.3\linewidth, trim={2mm 0 0 8mm}, clip = true]{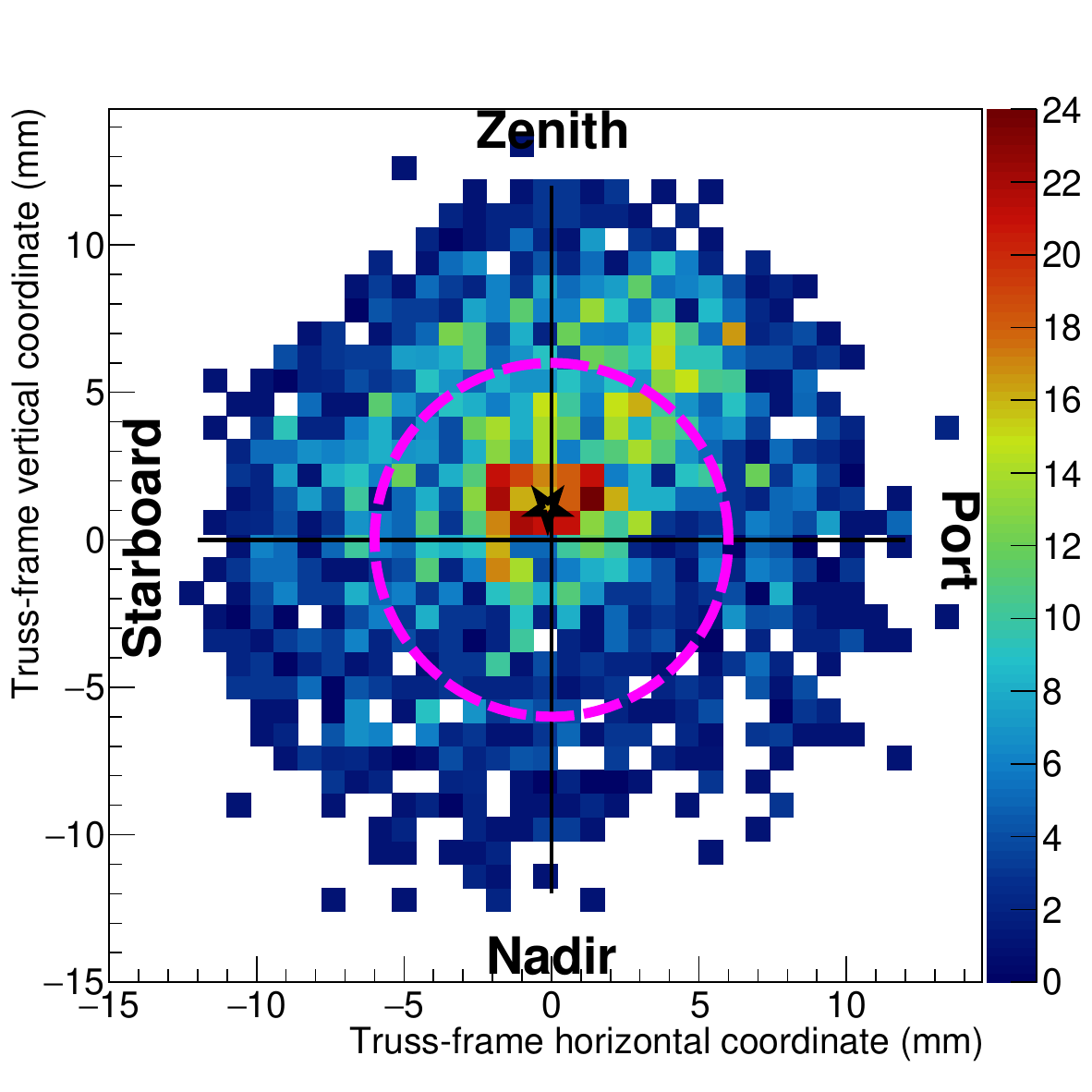}
\caption{
Imaging CZT images of Crab data from the three observations, July 11-13th, corresponding to Days 2-4, from top to bottom. Darker red bins correspond to higher photon detection rates. The pointing offset used as a center point for scattering and polarization analysis is given by the star shown on each image. This is calculated as a result of the weighted event distribution, taking into account the presence of inactive channels. Offsets of a few millimeters are typical during flight. A purple circle is shown at the location of the Beryllium scattering element. 
}
\label{fig:D17}

\end{figure}
Accurate reconstruction of polarization observables in \xlcal depends not only on precise event timing but also on a stable definition of the mirror focal point within the polarimeter aperture over the course of the flight. Small changes in pointing or optical alignment, if uncorrected, can introduce systematic distortions in the reconstructed azimuthal scatter-angle distribution and thus bias the inferred Stokes parameters. To mitigate these effects, \xlcal employs an imaging detector to track the effective optical axis on a day-by-day basis and to define a consistent reference frame for the polarization analysis.

This imaging detector is mounted directly behind the polarimeter and provides two-dimensional images of the focused X-ray beam, see~\citet{Abarr.2021} for more details on the instrument design. 
These images are generated on a day-by-day basis to track small shifts in the mirror alignment and to determine the true optical axis for each flight day as shown in Figure~\ref{fig:D17}.
The coordinate system for the images is centered on the central axis of the instrument. The image is then background-subtracted using data collected from off source pointing. Events are weighted based on the presence of dead detector channels. This is done to reduce bias in the offset measurement due to a non-uniform distribution of active channels. These event locations are then used to calculate a weighted average of the event location, giving a mean scattering site, from events within the diameter of the scattering element.
The resulting offsets, found in Table~\ref{table:offsets}, reflect changes in the gondola altitude, thermal flexing between the optics and detector assemblies, or slow drifts in the mirror alignment.
\begin{table}[H]
     \caption{
Calculated Mean Vertical and Mean Horizontal pointing offsets for each day of Crab Observation. These are derived by a weighted average of events impacting the imaging detector.}

     \centering
     \begin{tabular}{ccc}
         \toprule
         Observation & Vertical Offset [mm] & Horizontal Offset [mm]\\
         \midrule
        Day 2 &$1.41$& $0.32$\\

         Day 3 &$0.63$& $0.12$\\

         Day 4 &$1.08$&$ 0.13$\\
         \midrule
         Total &$0.99$&$ 0.21$\\
         \bottomrule
     \end{tabular}
\label{table:offsets}
 \end{table}

\clearpage
\section{Appendix B}
\counterwithin{figure}{section}
\label{App B}
MCMC corner plots depicting the probability distributions of model parameters for the phase correction of Day 4 and Day 2 respectively. The diagonal panels show the one-dimensional distributions for each individual parameter, highlighting the mode and 1$\sigma$ confidence interval, shown by dashed vertical lines. Off-diagonal panels display the joint distributions between parameter pairs, revealing correlations (e.g., changing \nutilde necessitates an appropriate shift in $\phi_5$), degeneracies (e.g., shifting the pulse profile by one bin width also returning reasonable fit results), or independence between them. The contours in off-diagonal plots represent 1, 2, and 3$\sigma$ confidence regions on a projection of the parameter space onto two-parameter planes. 
The histograms along the diagonal represent the marginalized distributions of $\phi_i$, \nutilde, and \nudottilde.
\begin{figure}[H]
    \centering
    \includegraphics[width=0.5\textwidth]{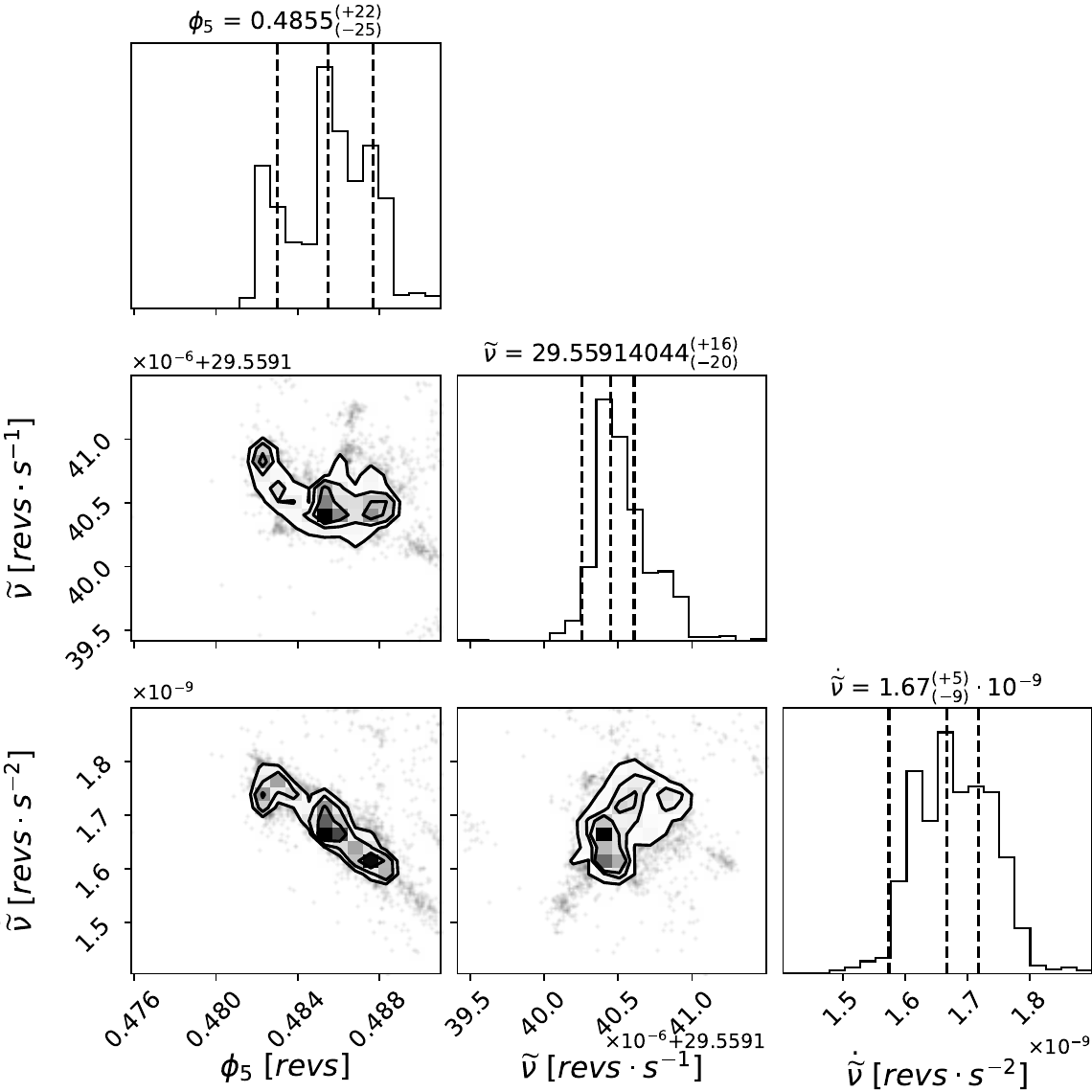}
    \caption{MCMC corner plot showing the posterior distributions of timing parameters used for correcting Day 4, the longest GPS-off interval, Interval 5. The diagonal panels show marginalized histograms for phase offset $\phi_5$, spin frequency \nutilde, and spin-down rate \nudottilde. Off-diagonal panels show the joint distributions with $1 \sigma$, $2 \sigma$, and $3 \sigma$ contours. On the diagonal, the central line depicts the mode of the distribution while $1 \sigma$ uncertainties are shown by the outer lines.}
    \label{fig:Day4MCMC}
\end{figure}
\clearpage

\begin{figure}[H]
    \centering
    \includegraphics[width=0.90\linewidth]{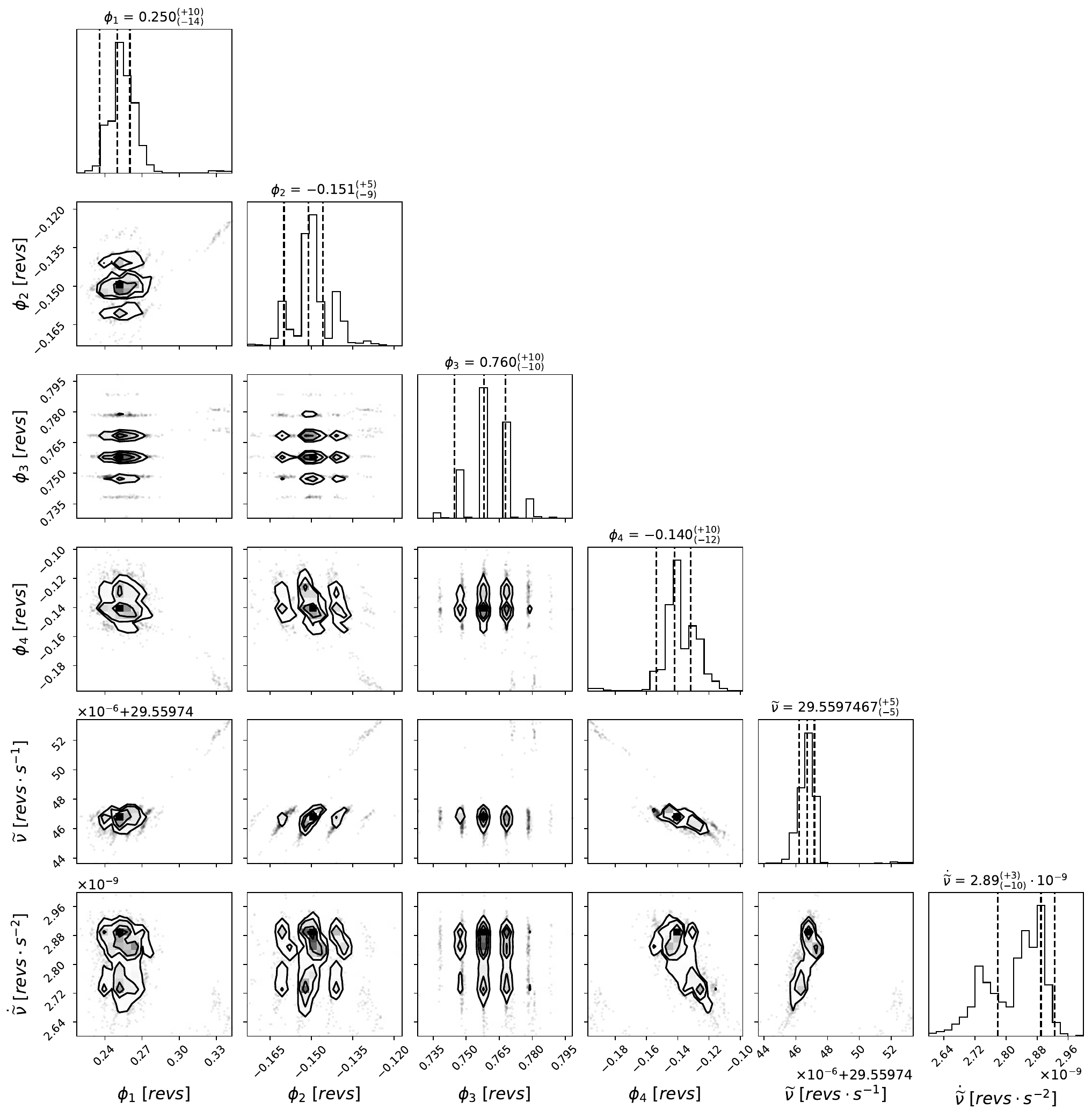}
    \caption{MCMC corner plot showing the probability distributions of timing parameters used for correcting Day 2. The diagonal panels depict 2D-histograms, illustrating correlation between combinations of parameters, for each phase offset $\phi_i$, spin frequency \nutilde, and spin-down rate \nudottilde. The central line depicts the mode of the distribution while $1 \sigma$ uncertainties are shown by the outer lines. The off-diagonal panels show the joint distributions with $1 \sigma$, $2 \sigma$, and $3 \sigma$ contours. Bi-modality in parameters $\phi_{1-4}$, \nutilde, and \nudottilde is possible due to limited statistics. Additionally, the stability of the on-board PPS signal likely depends on temperature, which would appear as multi-modal distribution in \nudottilde, due to the discontinuous nature of the Day 2 correction intervals.}
    \label{fig:Day2MCMC}
\end{figure}



\bibliography{crab}{}

@ARTICLE{2025MNRAS.540L..34A,
       author = {{Awaki}, Hisamitsu and {Baring}, Matthew G. and {Bose}, Richard and {Braun}, Dana and {Casey}, Jacob and {Chun}, Sohee and {Galchenko}, Pavel and {Gau}, Ephraim and {Goya}, Kazuho and {Hakamata}, Tomohiro and {Hayashi}, Takayuki and {Heatwole}, Scott and {Hu}, Kun and {Imazawa}, Ryo and {Ishi}, Daiki and {Ishida}, Manabu and {Kislat}, Fabian and {Kiss}, M{\'o}zsi and {Klepper}, Kassi and {Krawczynski}, Henric and {Kuramoto}, Haruki and {Lanzi}, R. James and {Lisalda}, Lindsey and {Maeda}, Yoshitomo and {af Malmborg}, Filip and {Matsumoto}, Hironori and {Menon}, Shravan Vengalil and {Miyamoto}, Aiko and {Miyamoto}, Asca and {Miyazawa}, Takuya and {Murakami}, Kaito and {Nagao}, Azuki and {Okajima}, Takashi and {Pearce}, Mark and {Rauch}, Brian F. and {Rodriguez Cavero}, Nicole and {Shima}, Kohei and {Shirahama}, Kentaro and {Snow}, Carlton M. and {Spooner}, Sean and {Takahashi}, Hiromitsu and {Takatsuka}, Sayana and {Tamura}, Keisuke and {Tanaka}, Kojiro and {Uchida}, Yuusuke and {West}, Andrew Thomas and {Wulf}, Eric A. and {Yokota}, Masato and {Yoshimoto}, Marina},
        title = "{XL-Calibur measurements of polarized hard X-ray emission from the Crab}",
      journal = {\mnras},
         year = 2025,
        month = jun,
       volume = {540},
       number = {1},
        pages = {L34-L40},
          doi = {10.1093/mnrasl/slaf026},
archivePrefix = {arXiv},
       eprint = {2503.14307},
 primaryClass = {astro-ph.HE},
       adsurl = {https://ui.adsabs.harvard.edu/abs/2025MNRAS.540L..34A}
}

@ARTICLE{1989A&A...221..180D,
       author = {{de Jager}, O.~C. and {Raubenheimer}, B.~C. and {Swanepoel}, J.~W.~H.},
        title = "{A powerful test for weak periodic signals with unknown light curve shape in sparse data.}",
      journal = {\aap},
         year = 1989,
        month = aug,
       volume = {221},
        pages = {180-190},
       adsurl = {https://ui.adsabs.harvard.edu/abs/1989A&A...221..180D}
}

@ARTICLE{2013PASP..125..306F,
       author = {{Foreman-Mackey}, Daniel and {Hogg}, David W. and {Lang}, Dustin and {Goodman}, Jonathan},
        title = "{emcee: The MCMC Hammer}",
      journal = {\pasp},
         year = 2013,
        month = mar,
       volume = {125},
       number = {925},
        pages = {306},
          doi = {10.1086/670067},
archivePrefix = {arXiv},
       eprint = {1202.3665},
 primaryClass = {astro-ph.IM},
       adsurl = {https://ui.adsabs.harvard.edu/abs/2013PASP..125..306F}
}

@ARTICLE{2019ApJ...871...12T,
       author = {{Timokhin}, A.~N. and {Harding}, A.~K.},
        title = "{On the Maximum Pair Multiplicity of Pulsar Cascades}",
      journal = {\apj},
         year = 2019,
        month = jan,
       volume = {871},
       number = {1},
          eid = {12},
        pages = {12},
          doi = {10.3847/1538-4357/aaf050},
archivePrefix = {arXiv},
       eprint = {1803.08924},
 primaryClass = {astro-ph.HE},
       adsurl = {https://ui.adsabs.harvard.edu/abs/2019ApJ...871...12T}
}

@ARTICLE{Kalapotharakos_2012,
    doi = {10.1088/0004-637X/749/1/2},
    url = {https://dx.doi.org/10.1088/0004-637X/749/1/2},
    year = {2012},
    month = {mar},
    publisher = {The American Astronomical Society},
    volume = {749},
    number = {1},
    pages = {2},
    author = {Kalapotharakos, Constantinos and Kazanas, Demosthenes and Harding, Alice and Contopoulos, Ioannis},
    title = {TOWARD A REALISTIC PULSAR MAGNETOSPHERE},
    journal = {The Astrophysical Journal}
}

@ARTICLE{2025A&A...693A.152G,
       author = {{Gonz{\'a}lez-Caniulef}, Denis and {Heyl}, Jeremy and {Fabiani}, Sergio and {Soffitta}, Paolo and {Costa}, Enrico and {Bucciantini}, Niccol{\`o} and {Kirmizibayrak}, Demet and {Xie}, Fei},
        title = "{Crab pulsar: IXPE observations reveal unified polarization properties in the optical and soft X-ray bands}",
      journal = {\aap},
         year = 2025,
        month = jan,
       volume = {693},
          eid = {A152},
        pages = {A152},
          doi = {10.1051/0004-6361/202451815},
archivePrefix = {arXiv},
       eprint = {2408.03245},
 primaryClass = {astro-ph.HE},
       adsurl = {https://ui.adsabs.harvard.edu/abs/2025A&A...693A.152G}
}

@ARTICLE{2010MNRAS.402.1027H,
       author = {{Hobbs}, G. and {Lyne}, A.~G. and {Kramer}, M.},
        title = "{An analysis of the timing irregularities for 366 pulsars}",
      journal = {\mnras},
         year = 2010,
        month = feb,
       volume = {402},
       number = {2},
        pages = {1027-1048},
          doi = {10.1111/j.1365-2966.2009.15938.x},
archivePrefix = {arXiv},
       eprint = {0912.4537},
 primaryClass = {astro-ph.GA},
       adsurl = {https://ui.adsabs.harvard.edu/abs/2010MNRAS.402.1027H}
}

@ARTICLE{2024A&A...687A.154W,
       author = {{Wang}, J. and {Verbiest}, J.~P.~W. and {Shaifullah}, G.~M. and {Cognard}, I. and {Guillemot}, L. and {Janssen}, G.~H. and {Mickaliger}, M.~B. and {Possenti}, A. and {Theureau}, G.},
        title = "{Improving pulsar timing precision through superior time-of-arrival creation}",
      journal = {\aap},
         year = 2024,
        month = jul,
       volume = {687},
          eid = {A154},
        pages = {A154},
          doi = {10.1051/0004-6361/202449366},
archivePrefix = {arXiv},
       eprint = {2405.08629},
 primaryClass = {astro-ph.IM},
       adsurl = {https://ui.adsabs.harvard.edu/abs/2024A&A...687A.154W}
}

@ARTICLE{2024ApJ...973..172W,
       author = {{Wong}, Josephine and {Mizuno}, Tsunefumi and {Bucciantini}, Niccol{\'o} and {Romani}, Roger W. and {Yang}, Yi-Jung and {Liu}, Kuan and {Deng}, Wei and {Goya}, Kazuho and {Xie}, Fei and {Pilia}, Maura and {Kaaret}, Philip and {Weisskopf}, Martin C. and {Silvestri}, Stefano and {Ng}, C. -Y. and {Chen}, Chien-Ting and {Agudo}, Iv{\'a}n and {Antonelli}, Lucio A. and {Bachetti}, Matteo and {Baldini}, Luca and {Baumgartner}, Wayne H. and {Bellazzini}, Ronaldo and {Bianchi}, Stefano and {Bongiorno}, Stephen D. and {Bonino}, Raffaella and {Brez}, Alessandro and {Capitanio}, Fiamma and {Castellano}, Simone and {Cavazzuti}, Elisabetta and {Ciprini}, Stefano and {Costa}, Enrico and {De Rosa}, Alessandra and {Del Monte}, Ettore and {Di Gesu}, Laura and {Di Lalla}, Niccol{\'o} and {Di Marco}, Alessandro and {Donnarumma}, Immacolata and {Doroshenko}, Victor and {Dov{\v{c}}iak}, Michal and {Ehlert}, Steven R. and {Enoto}, Teruaki and {Evangelista}, Yuri and {Fabiani}, Sergio and {Ferrazzoli}, Riccardo and {Garcia}, Javier A. and {Gunji}, Shuichi and {Heyl}, Jeremy and {Iwakiri}, Wataru and {Jorstad}, Svetlana G. and {Karas}, Vladimir and {Kislat}, Fabian and {Kitaguchi}, Takao and {Kolodziejczak}, Jeffery J. and {Krawczynski}, Henric and {La Monaca}, Fabio and {Latronico}, Luca and {Liodakis}, Ioannis and {Maldera}, Simone and {Manfreda}, Alberto and {Marin}, Fr{\'e}d{\'e}ric and {Marinucci}, Andrea and {Marscher}, Alan P. and {Marshall}, Herman L. and {Massaro}, Francesco and {Matt}, Giorgio and {Mitsuishi}, Ikuyuki and {Muleri}, Fabio and {Negro}, Michela and {O'Dell}, Stephen L. and {Omodei}, Nicola and {Oppedisano}, Chiara and {Papitto}, Alessandro and {Pavlov}, George G. and {Peirson}, Abel Lawrence and {Perri}, Matteo and {Pesce-Rollins}, Melissa and {Petrucci}, Pierre-Olivier and {Possenti}, Andrea and {Poutanen}, Juri and {Puccetti}, Simonetta and {Ramsey}, Brian D. and {Rankin}, John and {Ratheesh}, Ajay and {Roberts}, Oliver J. and {Sgr{\'o}}, Carmelo and {Slane}, Patrick and {Soffitta}, Paolo and {Spandre}, Gloria and {Swartz}, Douglas A. and {Tamagawa}, Toru and {Tavecchio}, Fabrizio and {Taverna}, Roberto and {Tawara}, Yuzuru and {Tennant}, Allyn F. and {Thomas}, Nicholas E. and {Tombesi}, Francesco and {Trois}, Alessio and {Tsygankov}, Sergey and {Turolla}, Roberto and {Vink}, Jacco and {Wu}, Kinwah and {Zane}, Silvia},
        title = "{Analysis of Crab X-Ray Polarization Using Deeper Imaging X-Ray Polarimetry Explorer Observations}",
      journal = {\apj},
         year = 2024,
        month = oct,
       volume = {973},
       number = {2},
          eid = {172},
        pages = {172},
          doi = {10.3847/1538-4357/ad6309},
archivePrefix = {arXiv},
       eprint = {2407.12779},
 primaryClass = {astro-ph.HE},
       adsurl = {https://ui.adsabs.harvard.edu/abs/2024ApJ...973..172W}
}

@article{KISLAT201545,
title = {Analyzing the data from X-ray polarimeters with Stokes parameters},
journal = {Astroparticle Physics},
volume = {68},
pages = {45-51},
year = {2015},
issn = {0927-6505},
doi = {https://doi.org/10.1016/j.astropartphys.2015.02.007},
url = {https://www.sciencedirect.com/science/article/pii/S092765051500033X},
author = {F. Kislat and B. Clark and M. Beilicke and H. Krawczynski}
}

@ARTICLE{2025arXiv250404775L,
       author = {{Li}, Hong and {Zhao}, Qing-Chang and {Feng}, Hua and {Tao}, Lian and {Tsygankov}, Sergey S.},
        title = "{Low-Count X-ray Polarimetry using the Bayesian Approach Reveals Fast Polarization Angle Variations}",
      journal = {arXiv e-prints},
         year = 2025,
        month = apr,
          eid = {arXiv:2504.04775},
        pages = {arXiv:2504.04775},
          doi = {10.48550/arXiv.2504.04775},
archivePrefix = {arXiv},
       eprint = {2504.04775},
 primaryClass = {astro-ph.HE},
       adsurl = {https://ui.adsabs.harvard.edu/abs/2025arXiv250404775L}
}

@ARTICLE{2018A&A...615A..54M,
       author = {{Mikhalev}, V.},
        title = "{Pitfalls of statistics-limited X-ray polarization analysis}",
      journal = {\aap},
         year = 2018,
        month = jul,
       volume = {615},
          eid = {A54},
        pages = {A54},
          doi = {10.1051/0004-6361/201731971},
archivePrefix = {arXiv},
       eprint = {1803.10120},
 primaryClass = {astro-ph.IM},
       adsurl = {https://ui.adsabs.harvard.edu/abs/2018A&A...615A..54M}
}

@ARTICLE{10.1093/mnras/stu2118,
    author = {Lyne, A. G. and Jordan, C. A. and Graham-Smith, F. and Espinoza, C. M. and Stappers, B. W. and Weltevrede, P.},
    title = {45 years of rotation of the Crab pulsar},
    journal = {Monthly Notices of the Royal Astronomical Society},
    volume = {446},
    number = {1},
    pages = {857-864},
    year = {2014},
    month = {11},
    issn = {0035-8711},
    doi = {10.1093/mnras/stu2118},
    url = {https://doi.org/10.1093/mnras/stu2118},
    eprint = {https://academic.oup.com/mnras/article-pdf/446/1/857/4154426/stu2118.pdf},
}

@ARTICLE{2018NatAs...2...50V,
       author = {{Vadawale}, S.~V. and {Chattopadhyay}, T. and {Mithun}, N.~P.~S. and {Rao}, A.~R. and {Bhattacharya}, D. and {Vibhute}, A. and {Bhalerao}, V.~B. and {Dewangan}, G.~C. and {Misra}, R. and {Paul}, B. and {Basu}, A. and {Joshi}, B.~C. and {Sreekumar}, S. and {Samuel}, E. and {Priya}, P. and {Vinod}, P. and {Seetha}, S.},
        title = "{Phase-resolved X-ray polarimetry of the Crab pulsar with the AstroSat CZT Imager}",
      journal = {Nature Astronomy},
         year = 2018,
        month = nov,
       volume = {2},
        pages = {50-55},
          doi = {10.1038/s41550-017-0293-z},
       adsurl = {https://ui.adsabs.harvard.edu/abs/2018NatAs...2...50V}
}

@article{Weisskopf.2000, 
year = {2000}, 
title = {{Discovery of Spatial and Spectral Structure in the X-Ray Emission from the Crab Nebula}}, 
author = {Weisskopf, Martin C. and Hester, J. Jeff and Tennant, Allyn F. and Elsner, Ronald F. and Schulz, Norbert S. and Marshall, Herman L. and Karovska, Margarita and Nichols, Joy S. and Swartz, Douglas A. and Kolodziejczak, Jeffery J. and O’Dell, Stephen L.}, 
journal = {The Astrophysical Journal Letters}, 
issn = {0004-637X}, 
doi = {10.1086/312733}, 
pmid = {10859123}, 
eprint = {astro-ph/0003216}, 
url = {http://arxiv.org/pdf/astro-ph/0003216.pdf}, 
pages = {L81--L84}, 
number = {2}, 
volume = {536}
}

@article{Chauvin.2018c7, 
year = {2018}, 
title = {{The PoGO+ view on Crab off-pulse hard X-ray polarisation}}, 
author = {Chauvin, M and Florén, H-G and Friis, M and Jackson, M and Kamae, T and Kataoka, J and Kawano, T and Kiss, M and Mikhalev, V and Mizuno, T and Tajima, H and Takahashi, H and Uchida, N and Pearce, M}, 
journal = {Monthly Notices of the Royal Astronomical Society: Letters}, 
issn = {1745-3925}, 
doi = {10.1093/mnrasl/sly027}, 
eprint = {1802.07775}
}

@article{Chauvin.2017, 
year = {2017}, 
title = {{Shedding new light on the Crab with polarized X-rays}}, 
author = {Chauvin, M. and Florén, H.-G. and Friis, M. and Jackson, M. and Kamae, T. and Kataoka, J. and Kawano, T. and Kiss, M. and Mikhalev, V. and Mizuno, T. and Ohashi, N. and Stana, T. and Tajima, H. and Takahashi, H. and Uchida, N. and Pearce, M.}, 
journal = {Scientific Reports}, 
doi = {10.1038/s41598-017-07390-7}, 
pmid = {28798398}, 
pmcid = {PMC5552847}, 
pages = {7816}, 
number = {1}, 
volume = {7}
}

@article{Bucciantini.2023, 
year = {2023}, 
title = {{Simultaneous space and phase resolved X-ray polarimetry of the Crab pulsar and nebula}}, 
author = {Bucciantini, Niccolò and Ferrazzoli, Riccardo and Bachetti, Matteo and Rankin, John and Lalla, Niccolò Di and Sgrò, Carmelo and Omodei, Nicola and Kitaguchi, Takao and Mizuno, Tsunefumi and Gunji, Shuichi and Watanabe, Eri and Baldini, Luca and Slane, Patrick and Weisskopf, Martin C. and Romani, Roger W. and Possenti, Andrea and Marshall, Herman L. and Silvestri, Stefano and Pacciani, Luigi and Negro, Michela and Muleri, Fabio and Wilhelmi, Emma de Oña and Xie, Fei and Heyl, Jeremy and Pesce-Rollins, Melissa and Wong, Josephine and Pilia, Maura and Agudo, Iván and Antonelli, Lucio A. and Baumgartner, Wayne H. and Bellazzini, Ronaldo and Bianchi, Stefano and Bongiorno, Stephen D. and Bonino, Raffaella and Brez, Alessandro and Capitanio, Fiamma and Castellano, Simone and Cavazzuti, Elisabetta and Chen, Chien-Ting and Ciprini, Stefano and Costa, Enrico and Rosa, Alessandra De and Monte, Ettore Del and Gesu, Laura Di and Marco, Alessandro Di and Donnarumma, Immacolata and Doroshenko, Victor and Dovčiak, Michal and Ehlert, Steven R. and Enoto, Teruaki and Evangelista, Yuri and Fabiani, Sergio and Garcia, Javier A. and Hayashida, Kiyoshi and Iwakiri, Wataru and Jorstad, Svetlana G. and Kaaret, Philip and Karas, Vladimir and Kislat, Fabian and Kolodziejczak, Jeffery J. and Krawczynski, Henric and Monaca, Fabio La and Latronico, Luca and Liodakis, Ioannis and Maldera, Simone and Manfreda, Alberto and Marin, Frédéric and Marinucci, Andrea and Marscher, Alan P. and Massaro, Francesco and Matt, Giorgio and Mitsuishi, Ikuyuki and Ng, C.-Y. and O’Dell, Stephen L. and Oppedisano, Chiara and Papitto, Alessandro and Pavlov, George G. and Peirson, Abel L. and Perri, Matteo and Petrucci, Pierre-Olivier and Poutanen, Juri and Puccetti, Simonetta and Ramsey, Brian D. and Ratheesh, Ajay and Roberts, Oliver J. and Soffitta, Paolo and Spandre, Gloria and Swartz, Doug and Tamagawa, Toru and Tavecchio, Fabrizio and Taverna, Roberto and Tawara, Yuzuru and Tennant, Allyn F. and Thomas, Nicolas E. and Tombesi, Francesco and Trois, Alessio and Tsygankov, Sergey and Turolla, Roberto and Vink, Jacco and Wu, Kinwah and Zane, Silvia}, 
journal = {Nature Astronomy}, 
doi = {10.1038/s41550-023-01936-8}, 
pages = {602--610}, 
number = {5}, 
volume = {7}
}

@article{Wong.2024, 
year = {2024}, 
title = {{Analysis of Crab X-ray Polarization using Deeper IXPE Observations}}, 
author = {Wong, Josephine and Mizuno, Tsunefumi and Bucciantini, Niccoló and Romani, Roger W and Yang, Yi-Jung and Liu, Kuan and Deng, Wei and Goya, Kazuho and Xie, Fei and Pilia, Maura and Kaaret, Philip and Weisskopf, Martin C and Silvestri, Stefano and Ng, C -Y and Chen, Chien-Ting and Agudo, Iván and Antonelli, Lucio A and Bachetti, Matteo and Baldini, Luca and Baumgartner, Wayne H and Bellazzini, Ronaldo and Bianchi, Stefano and Bongiorno, Stephen D and Bonino, Raffaella and Brez, Alessandro and Capitanio, Fiamma and Castellano, Simone and Cavazzuti, Elisabetta and Ciprini, Stefano and Costa, Enrico and Rosa, Alessandra De and Monte, Ettore Del and Gesu, Laura Di and Lalla, Niccoló Di and Marco, Alessandro Di and Donnarumma, Immacolata and Doroshenko, Victor and Dovčiak, Michal and Ehlert, Steven R and Enoto, Teruaki and Evangelista, Yuri and Fabiani, Sergio and Ferrazzoli, Riccardo and Garcia, Javier A and Gunji, Shuichi and Heyl, Jeremy and Iwakiri, Wataru and Jorstad, Svetlana G and Karas, Vladimir and Kislat, Fabian and Kitaguchi, Takao and Kolodziejczak, Jeffery J and Krawczynski, Henric and Monaca, Fabio La and Latronico, Luca and Liodakis, Ioannis and Maldera, Simone and Manfreda, Alberto and Marin, Frédéric and Marinucci, Andrea and Marscher, Alan P and Marshall, Herman L and Massaro, Francesco and Matt, Giorgio and Mitsuishi, Ikuyuki and Muleri, Fabio and Negro, Michela and O'Dell, Stephen L and Omodei, Nicola and Oppedisano, Chiara and Papitto, Alessandro and Pavlov, George G and Peirson, Abel Lawrence and Perri, Matteo and Pesce-Rollins, Melissa and Petrucci, Pierre-Olivier and Possenti, Andrea and Poutanen, Juri and Puccetti, Simonetta and Ramsey, Brian D and Rankin, John and Ratheesh, Ajay and Roberts, Oliver J and Sgró, Carmelo and Slane, Patrick and Soffitta, Paolo and Spandre, Gloria and Swartz, Douglas A and Tamagawa, Toru and Tavecchio, Fabrizio and Taverna, Roberto and Tawara, Yuzuru and Tennant, Allyn F and Thomas, Nicholas E and Tombesi, Francesco and Trois, Alessio and Tsygankov, Sergey and Turolla, Roberto and Vink, Jacco and Wu, Kinwah and Zane, Silvia}, 
journal = {arXiv}, 
doi = {10.48550/arxiv.2407.12779}, 
eprint = {2407.12779}
}

@article{Abarr.2021, 
year = {2021}, 
title = {{XL-Calibur – a second-generation balloon-borne hard X-ray polarimetry mission}}, 
author = {Abarr, Q. and Awaki, H. and Baring, M.G. and Bose, R. and Geronimo, G. De and Dowkontt, P. and Errando, M. and Guarino, V. and Hattori, K. and Hayashida, K. and Imazato, F. and Ishida, M. and Iyer, N.K. and Kislat, F. and Kiss, M. and Kitaguchi, T. and Krawczynski, H. and Lisalda, L. and Matake, H. and Maeda, Y. and Matsumoto, H. and Mineta, T. and Miyazawa, T. and Mizuno, T. and Okajima, T. and Pearce, M. and Rauch, B.F. and Ryde, F. and Shreves, C. and Spooner, S. and Stana, T.-A. and Takahashi, H. and Takeo, M. and Tamagawa, T. and Tamura, K. and Tsunemi, H. and Uchida, N. and Uchida, Y. and West, A.T. and Wulf, E.A. and Yamamoto, R.}, 
journal = {Astroparticle Physics}, 
issn = {0927-6505}, 
doi = {10.1016/j.astropartphys.2020.102529}, 
eprint = {2010.10608}, 
pages = {102529}, 
volume = {126}
}

@ARTICLE{pogo+,
       author = {{Chauvin}, M. and {Flor{\'e}n}, H.-G. and {Friis}, M. and {Jackson}, M. and {Kamae}, T. and {Kataoka}, J. and {Kawano}, T. and {Kiss}, M. and {Mikhalev}, V. and {Mizuno}, T. and {Tajima}, H. and {Takahashi}, H. and {Uchida}, N. and {Pearce}, M.},
        title = "{The PoGO+ view on Crab off-pulse hard X-ray polarization}",
      journal = {\mnras},
         year = 2018,
        month = jun,
       volume = {477},
       number = {1},
        pages = {L45-L49},
          doi = {10.1093/mnrasl/sly027},
archivePrefix = {arXiv},
       eprint = {1802.07775},
 primaryClass = {astro-ph.HE},
       adsurl = {https://ui.adsabs.harvard.edu/abs/2018MNRAS.477L..45C}
}

@ARTICLE{cyg,
       author = {{Awaki}, Hisamitsu and {Baring}, Matthew G. and {Bose}, Richard and {Casey}, Jacob and {Chun}, Sohee and {Dasgupta}, Adrika and {Galchenko}, Pavel and {Gau}, Ephraim and {Goya}, Kazuho and {Hakamata}, Tomohiro and {Hayashi}, Takayuki and {Heatwole}, Scott and {Hu}, Kun and {Ishi}, Daiki and {Ishida}, Manabu and {Kislat}, Fabian and {Kiss}, M{\'o}zsi and {Klepper}, Kassi and {Krawczynski}, Henric and {Kuramoto}, Haruki and {Lisalda}, Lindsey and {Maeda}, Yoshitomo and {Matsumoto}, Hironori and {Menon}, Shravan Vengalil and {Miyamoto}, Aiko and {Miyamoto}, Asca and {Murakami}, Kaito and {Okajima}, Takashi and {Pearce}, Mark and {Rauch}, Brian and {Rodriguez Cavero}, Nicole and {Shirahama}, Kentaro and {Spooner}, Sean and {Takahashi}, Hiromitsu and {Tamura}, Keisuke and {Uchida}, Yuusuke and {Wimalasena}, Kasun and {Yokota}, Masato and {Yoshimoto}, Marina},
        title = "{XL-Calibur Polarimetry of Cyg X-1 Further Constrains the Origin of Its Hard-state X-Ray Emission}",
      journal = {\apj},
         year = 2025,
        month = nov,
       volume = {994},
       number = {1},
          eid = {37},
        pages = {37},
          doi = {10.3847/1538-4357/ae0f1d},
archivePrefix = {arXiv},
       eprint = {2507.23126},
 primaryClass = {astro-ph.HE},
       adsurl = {https://ui.adsabs.harvard.edu/abs/2025ApJ...994...37A}
}

@article{ixpeCrab,
    author = {Wei, Wenhao and Xie, Fei and La Monaca, Fabio and Deng, Wei and Ge, Mingyu and Liu, Kuan and Zuo, Chao and Chen, Wei},
    title = {Energy-resolved polarization study of the Crab nebula with IXPE},
    journal = {Monthly Notices of the Royal Astronomical Society},
    volume = {539},
    number = {2},
    pages = {902-909},
    year = {2025},
    month = {04},
    issn = {0035-8711},
    doi = {10.1093/mnras/staf538},
    url = {https://doi.org/10.1093/mnras/staf538},
    eprint = {https://academic.oup.com/mnras/article-pdf/539/2/902/62922729/staf538.pdf},
}

@ARTICLE{1978ApJ...220L.117W,
       author = {{Weisskopf}, M.~C. and {Silver}, E.~H. and {Kestenbaum}, H.~L. and {Long}, K.~S. and {Novick}, R.},
        title = "{A precision measurement of the X-ray polarization of the Crab Nebula without pulsar contamination.}",
      journal = {\apjl},
         year = 1978,
        month = mar,
       volume = {220},
        pages = {L117-L121},
          doi = {10.1086/182648},
       adsurl = {https://ui.adsabs.harvard.edu/abs/1978ApJ...220L.117W}
}

@ARTICLE{1983ApJ...269..273W,
       author = {{Wilson}, R.~B. and {Fishman}, G.~J.},
        title = "{The pulse profile of the Crab pulsar in the energy range 45 keV-1.2 MeV.}",
      journal = {\apj},
         year = 1983,
        month = jun,
       volume = {269},
        pages = {273-280},
          doi = {10.1086/161039},
       adsurl = {https://ui.adsabs.harvard.edu/abs/1983ApJ...269..273W}
}

@ARTICLE{pogo,
       author = {{Chauvin}, M. and {Flor{\'e}n}, H.-G. and {Friis}, M. and {Jackson}, M. and {Kamae}, T. and {Kataoka}, J. and {Kawano}, T. and {Kiss}, M. and {Mikhalev}, V. and {Mizuno}, T. and {Ohashi}, N. and {Stana}, T. and {Tajima}, H. and {Takahashi}, H. and {Uchida}, N. and {Pearce}, M.},
        title = "{Shedding new light on the Crab with polarized X-rays}",
      journal = {Scientific Reports},
         year = 2017,
        month = aug,
       volume = {7},
          eid = {7816},
        pages = {7816},
          doi = {10.1038/s41598-017-07390-7},
archivePrefix = {arXiv},
       eprint = {1706.09203},
 primaryClass = {astro-ph.HE},
       adsurl = {https://ui.adsabs.harvard.edu/abs/2017NatSR...7.7816C}
}

@ARTICLE{Weisskopf2006,
       author = {{Weisskopf}, M.~C. and {Elsner}, R.~F. and {Hanna}, D. and {Kaspi}, V.~M. and {O'Dell}, S.~L. and {Pavlov}, G.~G. and {Ramsey}, B.~D.},
        title = "{The prospects for X-ray polarimetry and its potential use for understanding neutron stars}",
      journal = {arXiv e-prints},
         year = 2006,
        month = nov,
          eid = {astro-ph/0611483},
        pages = {astro-ph/0611483},
          doi = {10.48550/arXiv.astro-ph/0611483},
archivePrefix = {arXiv},
       eprint = {astro-ph/0611483},
 primaryClass = {astro-ph},
       adsurl = {https://ui.adsabs.harvard.edu/abs/2006astro.ph.11483W}
}

@article{Ng_2004,
doi = {10.1086/380486},
url = {https://doi.org/10.1086/380486},
year = {2004},
month = {jan},
publisher = {},
volume = {601},
number = {1},
pages = {479},
author = {Ng, C.-Y. and Romani, Roger W.},
title = {Fitting Pulsar Wind Tori},
journal = {The Astrophysical Journal}
}

@ARTICLE{Harding-2017-ApJ,
       author = {{Harding}, Alice K. and {Kalapotharakos}, Constantinos},
        title = "{Multiwavelength Polarization of Rotation-powered Pulsars}",
      journal = {\apj},
         year = 2017,
        month = may,
       volume = {840},
       number = {2},
          eid = {73},
        pages = {73},
          doi = {10.3847/1538-4357/aa6ead},
archivePrefix = {arXiv},
       eprint = {1704.06183},
 primaryClass = {astro-ph.HE},
       adsurl = {https://ui.adsabs.harvard.edu/abs/2017ApJ...840...73H}
}

@ARTICLE{Petri-2013-MNRAS,
       author = {{P{\'e}tri}, J.},
        title = "{Phase-resolved polarization properties of the pulsar striped wind synchrotron emission}",
      journal = {\mnras},
         year = 2013,
        month = sep,
       volume = {434},
       number = {3},
        pages = {2636-2644},
          doi = {10.1093/mnras/stt1214},
archivePrefix = {arXiv},
       eprint = {1308.0973},
 primaryClass = {astro-ph.HE},
       adsurl = {https://ui.adsabs.harvard.edu/abs/2013MNRAS.434.2636P}
}

@ARTICLE{2023PASJ...75.1298M,
       author = {{Mizuno}, Tsunefumi and {Ohno}, Hiroshi and {Watanabe}, Eri and {Bucciantini}, Niccol{\`o} and {Gunji}, Shuichi and {Shibata}, Sinpei and {Slane}, Patrick and {Weisskopf}, Martin C.},
        title = "{Magnetic-field structure of the Crab pulsar wind nebula revealed with IXPE}",
      journal = {\pasj},
         year = 2023,
        month = dec,
       volume = {75},
       number = {6},
        pages = {1298-1310},
          doi = {10.1093/pasj/psad070},
archivePrefix = {arXiv},
       eprint = {2309.16154},
 primaryClass = {astro-ph.HE},
       adsurl = {https://ui.adsabs.harvard.edu/abs/2023PASJ...75.1298M}
}

@ARTICLE{2024APh...15802944A,
       author = {{Aoyagi}, M. and {Bose}, R.~G. and {Chun}, S. and {Gau}, E. and {Hu}, K. and {Ishiwata}, K. and {Iyer}, N.~K. and {Kislat}, F. and {Kiss}, M. and {Klepper}, K. and {Krawczynski}, H. and {Lisalda}, L. and {Maeda}, Y. and {Malmborg}, F. af and {Matsumoto}, H. and {Miyamoto}, A. and {Miyazawa}, T. and {Pearce}, M. and {Rauch}, B.~F. and {Rodriguez Cavero}, N. and {Spooner}, S. and {Takahashi}, H. and {Uchida}, Y. and {West}, A.~T. and {Wimalasena}, K. and {Yoshimoto}, M.},
        title = "{Systematic effects on a Compton polarimeter at the focus of an X-ray mirror}",
      journal = {Astroparticle Physics},
         year = 2024,
        month = jun,
       volume = {158},
          eid = {102944},
        pages = {102944},
          doi = {10.1016/j.astropartphys.2024.102944},
archivePrefix = {arXiv},
       eprint = {2402.15229},
 primaryClass = {astro-ph.IM},
       adsurl = {https://ui.adsabs.harvard.edu/abs/2024APh...15802944A}
}

@ARTICLE{2026arXiv260209886B,
       author = {{Bouchet}, Tristan and {Laurent}, Philippe and {Cangemi}, Floriane and {Rodriguez}, J{\'e}r{\^o}me},
        title = "{In-flight calibration of the INTEGRAL/IBIS Compton mode: Application to the Crab Nebula polarization}",
      journal = {arXiv e-prints},
         year = 2026,
        month = feb,
          eid = {arXiv:2602.09886},
        pages = {arXiv:2602.09886},
          doi = {10.48550/arXiv.2602.09886},
archivePrefix = {arXiv},
       eprint = {2602.09886},
 primaryClass = {astro-ph.HE},
       adsurl = {https://ui.adsabs.harvard.edu/abs/2026arXiv260209886B}
}

@ARTICLE{2004ApJ...605L.129R,
       author = {{Rots}, Arnold H. and {Jahoda}, Keith and {Lyne}, Andrew G.},
        title = "{Absolute Timing of the Crab Pulsar with the Rossi X-Ray Timing Explorer}",
      journal = {\apjl},
         year = 2004,
        month = apr,
       volume = {605},
       number = {2},
        pages = {L129-L132},
          doi = {10.1086/420842},
archivePrefix = {arXiv},
       eprint = {astro-ph/0403187},
 primaryClass = {astro-ph},
       adsurl = {https://ui.adsabs.harvard.edu/abs/2004ApJ...605L.129R}
}
\bibliographystyle{aasjournalv7}



\end{document}